\documentclass[useAMS,usenatbib]{mn2e}
\usepackage{graphicx}
\usepackage{txfonts}
\usepackage[dvips,usenames]{color}
\def\aj{AJ}%
\def\araa{ARA\&A}%
\def\apj{ApJ}%
\def\apjl{ApJ}%
\def\apjs{ApJS}%
\def\aap{A\&A}%
\def\aaps{A\&AS}%
\def\jrasc{JRASC}%
\def\mnras{MNRAS}%
\def\prc{Phys.~Rev.~C}%
\def\pasp{PASP}%
\def\solphys{Sol.~Phys.}%
\def\ssr{Space~Sci.~Rev.}%
\def\zap{ZAp}%
\def\nat{Nature}%
\newcommand {\reac}[6] {$\rm\,{}^{#2}\kern-0.8pt{#1}\,({#3}\,,{#4})
  \,{}^{#6}\kern-0.8pt{#5}\,$}

\newcommand{\logt}{\mbox{$\log(t/{\rm yr})$}}
\newcommand{\Msun}{\mbox{$M_{\odot}$}}

\newcommand{\sub}[1]{\mbox{$_{\rm #1}$}}

\newcommand{\Teff}{\mbox{$T\sub{eff}$}}

\newcommand{\beq}{\begin{equation}}
\newcommand{\eeq}{\end{equation}}
\newcommand{\beqa}{\begin{eqnarray}}
\newcommand{\eeqa}{\end{eqnarray}}
\newcommand{\benu}{\begin{enumerate}}
\newcommand{\eenu}{\end{enumerate}}
\newcommand{\bite}{\begin{itemize}}
\newcommand{\eite}{\end{itemize}}
\newcommand{\bdes}{\begin{description}}
\newcommand{\edes}{\end{description}}

\newcommand{\refsec}[1]{Section \protect\ref{#1}}
\newcommand{\comment}[1]{}

\title[PARSEC: tracks and isochrones]{
  {\sl PARSEC}: stellar tracks and isochrones with the \\
  {\rm\sl{PA}}dova \& T{\rm\sl{R}}ieste {\rm\sl{S}}tellar
  {\rm\sl{E}}volution {\rm\sl{C}}ode}

\author[Bressan et al.]{Alessandro Bressan$^1$, Paola Marigo$^2$,
  L\'eo Girardi$^{3}$, Bernardo Salasnich$^{3}$, \newauthor Claudia
  Dal Cero$^{4}$, Stefano Rubele$^{3}$, Ambra Nanni$^1$
  \\
  $^1$ SISSA, via Bonomea 265, I-34136 Trieste, Italy \\
  $^2$ Dipartimento di Fisica e Astronomia Galileo Galilei,
  Universit\`a di Padova, Vicolo dell'Osservatorio 3, I-35122 Padova, Italy \\
  $^3$ Osservatorio Astronomico di Padova, Vicolo dell'Osservatorio 5,
  I-35122 Padova, Italy \\
  $^4$ Liceo Scientifico Paritario Gymnasium Patavinum Sport,
  via P.G.Guarneri 22, I-35132 Padova, Italy \\
}

\begin{document}

\date{Received April 2012 / Accepted .....}

\pagerange{\pageref{firstpage}--\pageref{lastpage}} \pubyear{2012}

\maketitle

\label{firstpage}

\begin{abstract}
  We present the updated version of the code used to compute stellar
  evolutionary tracks in Padova.  It is the result of a thorough
  revision of the major input physics, together with the inclusion of
  the pre--main sequence phase, not present in our previous releases
  of stellar models.  Another innovative aspect is the possibility of
  promptly generating accurate opacity tables fully consistent with
  {\em any} selected initial chemical composition, by coupling the
  OPAL opacity data at high temperatures to the molecular opacities
  computed with our \AE SOPUS code \citep{MarigoAringer_09}.  In this
  work we present extended sets of stellar evolutionary models for
  various initial chemical compositions, while other sets with
  different metallicities and/or different distributions of heavy
  elements are being computed.  For the present release of models we
  adopt the solar distribution of heavy elements from the recent
  revision by \citet{Caffau_etal11}, corresponding to a Sun's
  metallicity $Z\!\simeq\!0.0152$.  From all computed sets of stellar
  tracks, we also derive isochrones in several photometric systems.
  The aim is to provide the community with the basic tools to model
  star clusters and galaxies by means of population synthesis
  techniques.
\end{abstract}

\begin{keywords}
  stars: evolution -- stars: interiors -- Hertz\-sprung--Russel (HR)
  diagram -- stars: low mass
\end{keywords}

\section{ Introduction }
\label{intro}

In this paper we briefly describe our new stellar evolution code {\bf
  PARSEC}: the {\bf{PA}}dova \& {\small{T}}{\bf{R}}ieste
{\bf{S}}tellar {\bf{E}}volution {\bf{C}}ode.  This is the result of a
thorough revision and update of the stellar evolution code used in
Padova to compute sets of stellar evolutionary tracks that are widely
used by the astronomical community \citep{Bressan_etal93,
  Bertelli_etal94, Girardi_etal00, Marigo_etal01, Bertelli_etal08,
  Bertelli_etal09}.

As we describe next, the most important changes include the updating
of the major input physics (equation of state, opacities, nuclear
reaction rates), the implementation of microscopic diffusion, and the
extension of the evolutionary calculations to the pre-main sequence
phase (PMS).

We have devoted much care to ensure consistency between the adopted
chemical mixture and the physical ingredients in {\em PARSEC}, i.e.
opacities and equation of state, as detailed in Sects.~\ref{sec_opac},
and \ref{sec_eos} respectively.  In this respect a distinctive
prerogative of our stellar models is the use of the \AE SOPUS tool
\citep{MarigoAringer_09}, that enables us to compute tables of
accurate low-$T$ opacities for {\sl any} specified mixture of chemical
elements from H to U, just starting from the monochromatic absorption
coefficients of all opacity sources under consideration (e.g.\ line
lists of several molecules).  We have put the most possible effort to
keep the same abundance flexibility also when dealing with the
equation of state, the high-temperature radiative opacities and the
conductive opacities, for which we have employed other public codes.
This represents a key aspect since we are now able to predict readily
the evolution of stars for any chemical pattern of interest (varying
CNO abundances, different degrees of enhancement/reduction in
$\alpha$-elements, C-N, Ne-O and Mg-Al abundance anti-correlations
suitable for Galactic globular clusters, etc.).

In this paper we present new sets of evolutionary tracks for the
initial chemical compositions in the range from $Z=0.0005$ to
$Z=0.07$. The helium content $Y$ is assumed to increase with $Z$.
Other metallicities and arbitrary distributions of heavy elements are
being considered.  The range of initial masses ranges from
$0.1~M_{\odot}$ to $12~M_{\odot}$, and the evolutionary phases extend
from the PMS till either the onset thermally-pulsing AGB phase
(TP-AGB) or carbon ignition. From all these tracks, we also derive
isochrones in several photometric systems.

The plan of this paper is as follows.  \refsec{sec_input} presents the
input physics of the models which includes the new solar distribution
of elements; \refsec{sec_sun} deals with the calibration of important
model parameters with the Solar model; \refsec{sec_chemic} briefly
describes other adopted element distributions and global
metallicities; \refsec{sec_tracks} introduces the new stellar tracks
and discuss their main characteristics and \refsec{sec_isochrones}
describes the corresponding isochrones.


\section{Input physics}
\label{sec_input}

\subsection{The solar distribution of heavy elements}
\label{distribution}

Before discussing any other relevant input physics we specify the {\em
  solar} distribution of heavy elements adopted in this paper. For
each element heavier than $^4$He, we must assign its fractional
abundance relative to the total solar metallicity, i.e.\
$X_{i,\odot}/Z_{\odot}$.  This will be the reference distribution, with
respect to which other {\em non-scaled solar} mixtures will be
considered.

The reference solar distribution of metals consists of $90$ chemical
elements\footnote{A few elements (Po, At, Rn, Fr, Ra, Ac, and Pa) are
  assigned negligible abundances.} from Li to U, with abundances taken
from the compilation by \citet{GrevesseSauval_98}, except for a subset
of species for which we adopt the recommended values according to the
latest revision by \citet{Caffau_etal11} and references therein.  The
solar abundances of the recently revised elements are listed in
Table~\ref{tab_solarcomp}\footnote{Some of the elements listed in
  Table~\ref{tab_solarcomp} (e.g. Eu, Hf, Th) may be irrelevant for
  the evolutionary calculations presented in this paper, however they
  will become relevant in the context of the TP-AGB tracks, to be
  described in subsequent papers.}.

According to this abundance compilation, the present-day Sun's
metallicity is $Z_{\odot}= 0.01524$, that can be compared to other
recent estimates, e.g.  $Z_{\odot}= 0.0141$ of \citet{Lodders_etal09},
and $Z_{\odot}= 0.0134$ of \citet{Asplund_etal09}.

In addition to the solar one, other chemical distributions have been
considered in this work, which should be representative of galaxies
with specific chemical evolution histories such as, for instance, the
Magellanic Clouds, massive early-type galaxies or other systems with
$\alpha$-enhanced or $\alpha$-depleted patterns.

\begin{table}
\caption{Solar abundances of a few elements adopted in this
work, following the values recommended by \citet{Caffau_etal11}
and references therein. Abundances are expressed with the standard
notation $A(Y) = \log (n_Y/n_{\rm H}) + 12$.
For all other species we adopt the compilation
of \citet{GrevesseSauval_98}.}
\label{tab_solarcomp}
\begin{tabular}{lcl}
\hline
\multicolumn{1}{c}{Element} & \multicolumn{1}{c}{Abundance $A(Y)$} & \multicolumn{1}{c}{Reference} \\
\hline
Li &  $1.03$ & \citet{Caffau_etal11}\\
C  &  $8.50$ & \citet{Caffau_etal10}\\
N  &  $7.86$ & \citet{Caffau_etal09}\\
O  &  $8.76$ & \citet{Caffau_etal08b}\\
P  &  $5.46$ & \citet{Caffau_etal07b}\\
S  &  $7.16$ & \citet{CaffauLudwig_07}\\
K  &  $5.11$ & \citet{Caffau_etal11}\\
Fe &  $7.52$ & \citet{Caffau_etal11}\\
Eu &  $0.52$ & \citet{Mucciarelli_etal08}\\
Hf &  $0.87$ & \citet{Caffau_etal08a}\\
Os &  $1.36$ & \citet{Caffau_etal11}\\
Th &  $0.08$ & \citet{Caffau_etal08a}\\
\hline
\end{tabular}
\end{table}

\begin{figure*}
\begin{minipage}{0.48\textwidth}
\resizebox{\hsize}{!}{\includegraphics{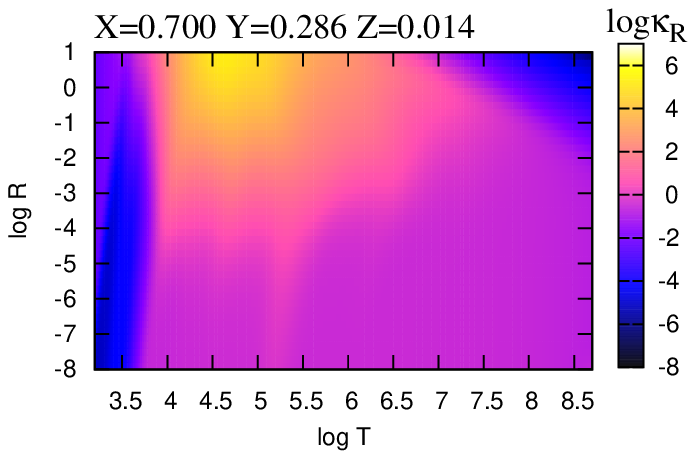}}
\end{minipage}
\hfill
\begin{minipage}{0.48\textwidth}
\resizebox{\hsize}{!}{\includegraphics{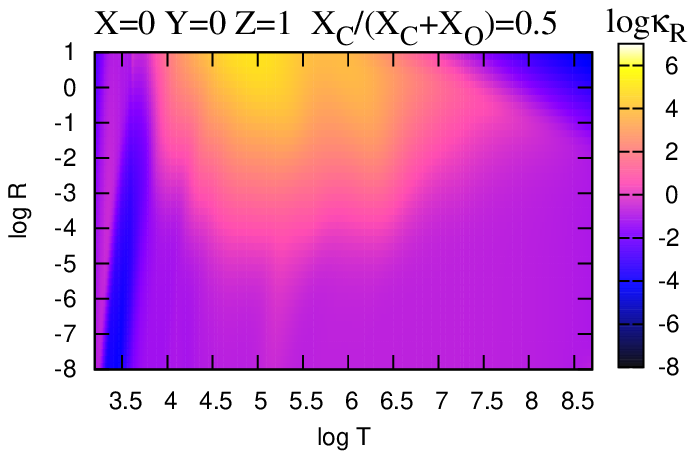}}
\end{minipage}
\caption{Maps of Rosseland mean opacities showing the actual coverage
of the tables in the $\log(T)-\log(R)$ plane. While both plots refer to the
same reference metallicity $Z=0.014$, the left panel exemplifies the case
of an ``H-rich'' table, the right panel corresponds to a ``H-free'' table,
suitable to describe the opacity of a gas in which
all helium has been burnt and converted
into carbon and oxygen in the same proportions.}
\label{fig_opatab}
\end{figure*}

\subsection{Opacities}
\label{sec_opac}
Following a standard procedure, in PARSEC we describe the absorption
properties of matter in the gas phase by means of pre-computed, static
tables of Rosseland mean opacities, $\kappa_{\rm R}(\rho,T)$, which
are suitably arranged to encompass a region of the temperature-density
($T-\rho$) plane wide enough to cover all values met across the
stellar structure during the evolution\footnote{Under the assumption
  of local thermodynamical equilibrium and in the limit of diffusion
  approximation, the solution of the radiation transfer equation
  simplifies in such a way that the flux of radiation $F$ as a
  function of radius $r$ is expressed as $F(r) \propto 1/\kappa_{\rm
    R}$.}.

Beside the state variables, $T$ and $\rho$, opacities depend on the
chemical composition of the gas, which is commonly specified by a set
of abundance parameters, i.e.: the total metallicity $Z$, the hydrogen
abundance $X$, and the distributions $\{X_i/Z\}$ of the heavy elements in
the mixture. The latter depend on the specific case under
consideration, e.g.  scaled-solar mixtures with $\{X_i/Z\} =
\{X_i/Z\}_{\odot}$, or others $\{X_i/Z\}$ derived from various
constraints such as the enhancement/depletion of $\alpha$-elements
(expressed by the ratio [$\alpha$/Fe]), or the over-abundances in
primary C and O necessary to describe the He-burning regions.  For the
computation of the opacity tables we employ different programs as
detailed below.

In the high-temperature regime, $4.2 \le \log(T/{\rm K}) \le 8.7$, we
adopt the opacity tables provided by the Opacity Project At Livermore
(OPAL) team \citep[][and references therein]{IglesiasRogers_96}. We
use the OPAL interactive web
mask\footnote{http://opalopacity.llnl.gov/} specifying the number
fractions of $19$ heavy elements (C, N, O, Ne, Na, Mg, Al, Si, P, S,
Cl, Ar, K, Ca, Ti, Cr, Mn, Fe, and Ni), as implied by our set of
${X_i/Z}$.

In the low-temperature regime, $3.2 \le \log(T/{\rm K}) \le 4.1$, we
employ the \AE SOPUS\footnote{http://stev.oapd.inaf.it/aesopus} tool
\citep{MarigoAringer_09} to generate opacity tables for any specified
set of chemical abundances for $92$ elements from H to U. \AE SOPUS
solves the equation of state of matter in the gas phase for $\approx
800$ chemical species consisting of almost $300$ atoms (neutral atoms
and ions up to $5^{\rm th}$ ionization stage) and $500$ molecular
species. The gas opacities account for many continuum and discrete
sources, including atomic opacities, molecular absorption, and
collision-induced absorption.

In the transition interval $4.0<\log(T/{\rm K})<4.1$, a linear
interpolation between the opacities derived from the OPAL and \AE
SOPUS is adopted.  We remind the reader that both opacities sources
provide values in good agreement in this temperature interval.  As
shown by \citet[][see their figure 7]{MarigoAringer_09} the
logarithmic differences between OPAL and \AE SOPUS opacities are
lower than $0.05$ dex for most cases.

Conductive opacities are included following \citet{Itoh_etal08}. In
the computation, for any specified chemical mixture, the total thermal
conductivity accounts for the contribution of $11$ atomic species
($^1$H, $^4$He, $^{12}$C, $^{14}$N, $^{16}$O, $^{20}$Ne, $^{24}$Mg,
$^{28}$Si, $^{32}$S, $^{40}$Ca, and $^{56}$Fe), each weighted by the
corresponding abundance (by number). To this aim we have implemented
in our code the fortran routine kindly made available by Itoh (private
communication).

In practice, given the total reference metallicity $Z$ and the
distribution of heavy elements $\{X_i/Z\}$, we construct two sets of
opacity tables, to which we simply refer to as ``H-rich'' and
``H-free'' opacities.  The former set comprises $N_X$ opacity tables,
where $N_X$ denotes the number (typically $10$) of hydrogen abundance
values, ranging from $X=0$ to $X=1-Z$.  The latter set is
characterised by $X=0$, while the helium content assumes $N_Y$ values
(typically $N_Y=10$), ranging from $Y=0$ to $Y=1-Z$.  In addition, for
each $Y$ value, we consider three combinations of C and O abundances,
defined by the ratios $R_{\rm C}=X_{\rm C}/(X_{\rm C}+X_{\rm O})=0.0,
0.5, 1.0$.  All other elements are left unchanged.  The ``H-free''
tables are specifically designed to describe the opacity in the
He-burning regions.

For both sets, each opacity table covers a rectangular region defined
by the intervals $3.2 \le \log(T/{\rm K}) \le 8.7$ and $-8 \le \log(R)
\le 1$, where $R = \rho/T_6$ in g\,cm$^{-3}$, and $T_6=T/(10^6\,{\rm
  K})$.  To limit as much as possible the accuracy loss due to
subsequent interpolation we adopt a fine grid spacing, with
$\Delta\log T = 0.01$ for $3.2 \le \log(T/{\rm K}) < 3.7$, $\Delta\log
T = 0.02$ for $3.7 \le \log(T/{\rm K}) \le 8.7$, and $\Delta\log R =
0.2$.  Two examples of opacity tables for both the ``H-rich'' and
``H-free'' cases are given in Fig.~\ref{fig_opatab}. The maps are
obtained with the same grid spacing of the tables, which actually
provides quite a dense and smooth description of the opacity all over
the thermodynamic domain of interest.

Interpolation over ``H-rich'' tables is perfomed in four dimensions,
i.e. using $R$, $T$, $X$, and $Z$ as the independent variables.  While
the interpolation in $R$ and $T$ is bilinear, we adopt a parabolic
scheme for both $X$ and $Z$ interpolation.  Interpolation over
``H-free'' tables is perfomed in five dimensions, i.e. involving $R$,
$T$, $Y$, $R_{\rm C}=X_{\rm C}/(X_{\rm C}+X_{\rm O})$, and $Z$.
Interpolation is bilinear in $R$ and $T$, linear in $R_{\rm C}$, while
we use as before a parabolic scheme for $Z$ interpolation.

Actually, bilinear interpolation in $R$ and $T$ does not preserve the
continuity of the first derivatives, possibly introducing some
instability into the convergence of the stellar model.  For this
reason, once the opacity tables are loaded, we compute and store the
logarithmic derivatives of the opacities (with respect to $T$ and $R$)
over the same grid of the opacity tables.  Hence, the partial
derivatives can be obtained by bilinear interpolation, in the same way
as for the opacity values.  While fulfilling the continuity
requirements of the derivatives, this method has also the advantage of
being quite fast.  It should be also specified that, for each planned
set of stellar tracks, the PARSEC code preliminarly loads a suitable
number $N_Z$ of opacity tables in order to follow in detail any
significant change in the local metal content $Z$ due, for instance,
to the diffusion of heavy elements or dredge-up episodes.

\subsection{Equation of state}
\label{sec_eos}
For the equation of state (EOS) we make use of the FreeEOS code
developed and updated over the years by A.W.~Irwin, and freely
available under the GPL
licence\footnote{http://freeeos.sourceforge.net/}.  The FreeEOS
package is fully implemented in our code and we may use it
``on-the-fly'', for different approximations and levels of accuracy.
However, since the pre-tabulated version is sufficiently accurate for
most of our purposes, we proceed by pre-computing suitable tables and
by interpolating between them.

The EOS calculation is performed accounting for the contributions of
several elements, namely: H, He, C, N, O, Ne, Na, Mg, Al, Si, P, S,
Cl, Ar, Ca, Ti, Cr, Mn, Fe, and Ni.  For any specified distribution of
heavy elements $\{X_i/Z\}$, we consider several values of the
metallicity $Z$, and for each value of $Z$ we pre-compute tables
cointaning all thermodynamic quantities of interest (e.g. mass
density, mean molecular weight, entropy, specific heats and their
derivatives, etc.)  over suitably wide ranges of temperature and
pressure.  Exactly in the same fashion as for the opacity, we arrange
a ``H-rich'' set containing $N_X=10$ tables each characterised by
different H abundances, and a ``H-free'' set consisting of $31$
tables, which are designed to describe He-burning regions.  In
practice we consider $10$ values of the helium abundance, from $Y=0$
to $Y=1-Z$. For each $Y$ we compute three tables with C and O
abundances determined by the ratios $R_{\rm C}=X_{\rm C}/(X_{\rm
  C}+X_{\rm O})=0.0, 0.5, 1.0$.

Multi-dimension interpolations (in the variables $Z$, $X$ or $Y$ and
$R_{\rm C}$) are carried out with the same scheme adopted for the
opacities.  All interesting derivatives are pre-computed and included
in the EOS tables.

\begin{figure*}
  \hfill
  \includegraphics[angle=90,width=0.45\textwidth]{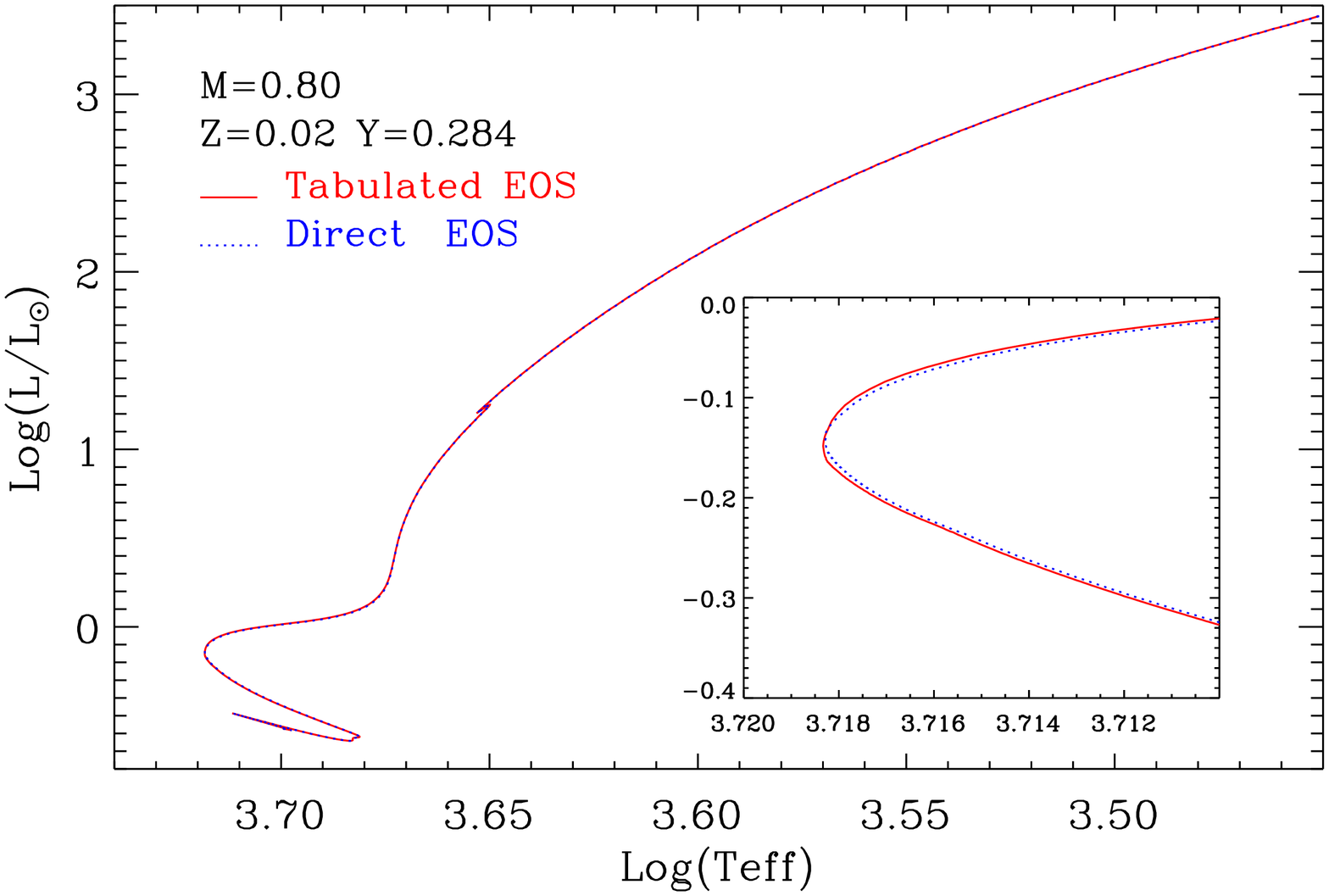}
  \hfill
  \includegraphics[angle=90,width=0.45\textwidth]{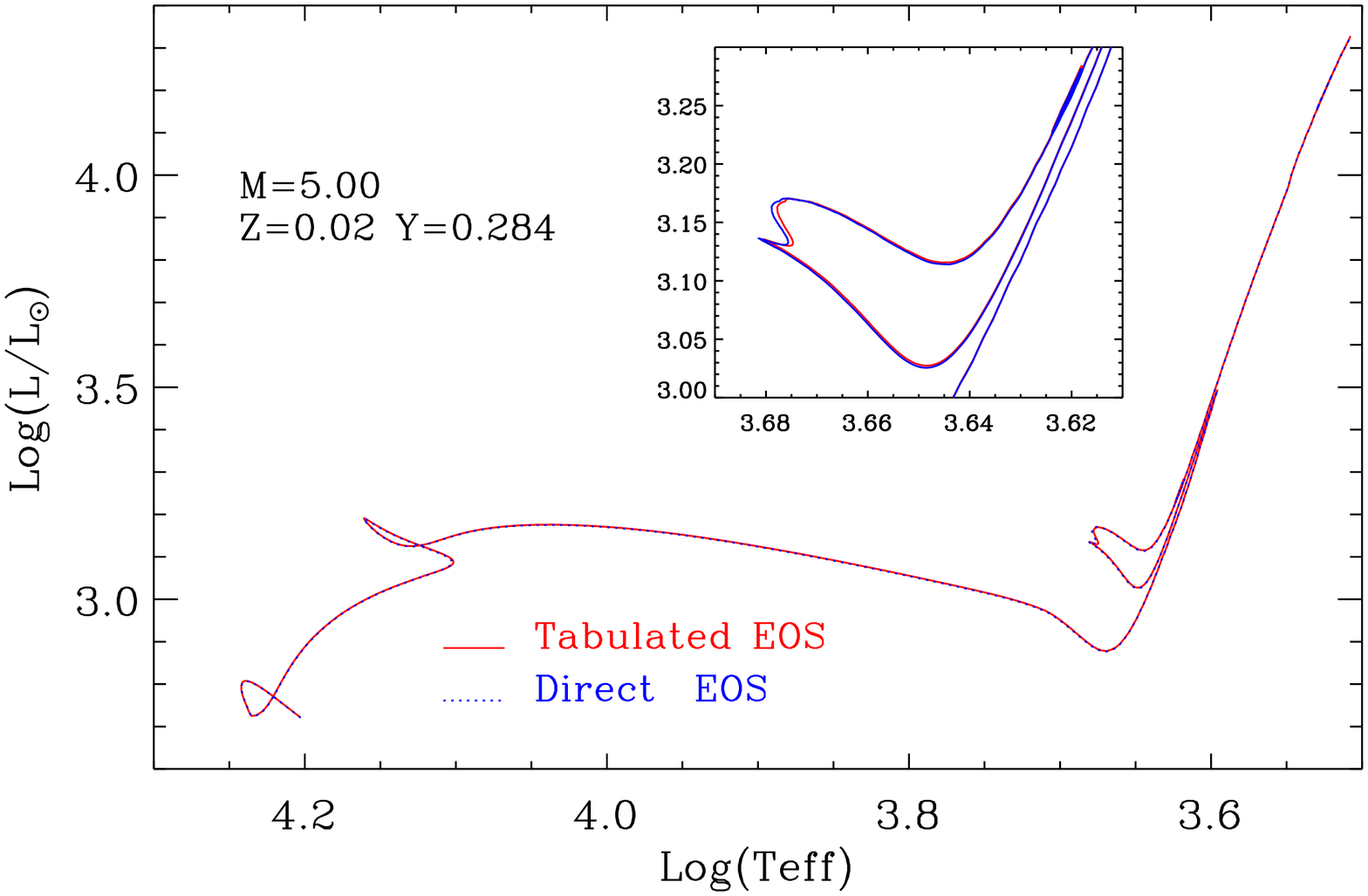}
  \caption{Comparison between evolutionary tracks computed with the
    direct calculation of FreeEOS, or interpolating between
    pre-tabulated FreeEOS tables.  The differences between these two
    cases are clearly negligible.}
    \label{fig_freeeos}
\end{figure*}

Our procedure is to minimize the effects of interpolation by computing
a set of EOS tables exactly with the initial metallicity and partition
of the new set of tracks.  This set is then inserted into the EOS
database for interpolation when the global metallicity $Z$ changes
during the evolution.

Before proceeding with the calculation of evolutionary tracks, we
extensively tested the accuracy of the interpolation method.
Fig.~\ref{fig_freeeos} illustrates the typical results from these
tests: it compares evolutionary tracks computed with either the
interpolation within the tabulated EOS, or a direct call to FreeEOS
for each mesh point. As can be appreciated, the differences in the HR
diagram are clearly negligible.  The same applies to other quantities
in the tracks, such as the lifetimes and the luminosity at the RGB
tip.

\subsection{Nuclear reaction rates}
\label{sec_rates}

Our nuclear network consists of the p-p chains, the CNO tri-cycle, the
Ne--Na and Mg--Al chains, and the most important $\alpha$-capture
reactions, including the $\alpha$-n reactions.  The network solves for
the abundances of $N_{\rm el} = 26$ chemical species: $^1$H, D,
$^3$He, $^4$He,$^7$Li, $^8$Be, $^4$He, $^{12}$C, $^{13}$C, $^{14}$N,
$^{15}$N, $^{16}$N, $^{17}$N, $^{18}$O, $^{19}$F, $^{20}$Ne,
$^{21}$Ne, $^{22}$Ne, $^{23}$Na, $^{24}$Mg, $^{25}$Mg, $^{26}$Mg,
$^{26}$Al$^m$, $^{26}$Al$^g$, $^{27}$Al, $^{28}$Si. The latter nucleus
acts as the ``exit element'', which terminates the network.  In total
we consider $42$ reaction rates, listed in Tab.~\ref{tab_rates}.
These are the recommended rates in the JINA reaclib database
\citep{Cyburt_etal10}, from which we also take the $Q$-value of each
reaction.

\begin{table}
\label{tab_rates}
 \centering
  \caption{Nuclear reaction rates adopted in this work.}
  \begin{tabular}{@{}ll@{}}
  \hline
\multicolumn{1}{c}{Reaction} &
\multicolumn{1}{c}{Reference} \\
\hline
\reac{p}{}{p}{\beta^+\,\nu}{D}{} & \citet{Cyburt_etal10}  \\
\reac{p}{}{D}{\gamma}{He}{3} &  \citet{Descouvemont_etal04}\\
\reac{He}{3}{^{3}He}{\gamma}{2\,p + ^{4}\kern-0.8pt{He}}{} &  \citet{Angulo_etal99}\\
\reac{He}{4}{^{3}He}{\gamma}{Be}{7} &  \citet{Descouvemont_etal04}\\
\reac{Be}{7}{e^-}{\gamma}{Li}{7} &  \citet{CaughlanFowler_88}\\
\reac{Li}{7}{p}{\gamma}{^{4}\kern-2.0pt{He} + ^{4}\kern-2.0pt{He}}{} &  \citet{Descouvemont_etal04} \\
\reac{Be}{7}{p}{\gamma}{B}{8} & \citet{Angulo_etal99}  \\
\reac{C}{12}{p}{\gamma}{N}{13} & \citet{Angulo_etal99}  \\
\reac{C}{13}{p}{\gamma}{N}{14} & \citet{Angulo_etal99}  \\
\reac{N}{14}{p}{\gamma}{O}{15} & \citet{Imbriani_etal05}  \\
\reac{N}{15}{p}{\gamma}{^4He + ^{12}\kern-2.0pt{C}}{} & \citet{Angulo_etal99}  \\
\reac{N}{15}{p}{\gamma}{O}{16} & \citet{Angulo_etal99}  \\
\reac{O}{16}{p}{\gamma}{F}{17} & \citet{Angulo_etal99}  \\
\reac{O}{17}{p}{\gamma}{\,^4He + ^{14}\kern-2.0pt{N}}{} & \citet{Chafa_etal07}  \\
\reac{O}{17}{p}{\gamma}{F}{18} & \citet{Chafa_etal07}  \\
\reac{O}{18}{p}{\gamma}{\,^4He + ^{15}\kern-2.0pt{N}}{} & \citet{Angulo_etal99}  \\
\reac{O}{18}{p}{\gamma}{F}{19} & \citet{Angulo_etal99}  \\
\reac{F}{19}{p}{\gamma}{\,^4He + ^{16}\kern-2.0pt{O}}{} & \citet{Angulo_etal99}  \\
\reac{F}{19}{p}{\gamma}{Ne}{20} & \citet{Angulo_etal99}  \\
\reac{Ne}{20}{p}{\gamma}{Na}{21} & \citet{Angulo_etal99}  \\
\reac{Ne}{21}{p}{\gamma}{Na}{22} & \citet{Iliadis_etal01}  \\
\reac{Ne}{22}{p}{\gamma}{Na}{23} & \citet{Hale_etal02}  \\
\reac{Na}{23}{p}{\gamma}{\,^4He + ^{20}\kern-2.0pt{Ne}}{} & \citet{Hale_etal04}  \\
\reac{Na}{23}{p}{\gamma}{Mg}{24} & \citet{Hale_etal04}  \\
\reac{Mg}{24}{p}{\gamma}{Al}{25} & \citet{Iliadis_etal01}  \\
\reac{Mg}{25}{p}{\gamma}{Al^g}{26} & \citet{Iliadis_etal01}  \\
\reac{Mg}{25}{p}{\gamma}{Al^m}{26} & \citet{Iliadis_etal01}  \\
\reac{Mg}{26}{p}{\gamma}{Al}{27} & \citet{Iliadis_etal01}  \\
\reac{Al^g}{26}{p}{\gamma}{Si}{27} & \citet{Iliadis_etal01}  \\
\reac{Al}{27}{p}{\gamma}{\,^4He + ^{24}\kern-2.0pt{Mg}}{} & \citet{Iliadis_etal01}  \\
\reac{Al}{27}{p}{\gamma}{Si}{28} & \citet{Iliadis_etal01}  \\
\reac{He}{4}{2\,^{4}He}{\gamma}{C}{12} & \citet{Fynbo_etal05}  \\
\reac{C}{12}{^{4}He}{\gamma}{O}{16} & \citet{Buchmann_96}  \\
\reac{N}{14}{^{4}He}{\gamma}{F}{18} & \citet{Gorres_etal00}  \\
\reac{N}{15}{^{4}He}{\gamma}{F}{19} & \citet{Wilmes_etal02}  \\
\reac{O}{16}{^{4}He}{\gamma}{Ne}{20} & \citet{Angulo_etal99}  \\
\reac{O}{18}{^{4}He}{\gamma}{Ne}{22} & \citet{Dababneh_etal03}  \\
\reac{Ne}{20}{^{4}He}{\gamma}{Mg}{24} & \citet{Angulo_etal99}  \\
\reac{Ne}{22}{^{4}He}{\gamma}{Mg}{26} & \citet{Angulo_etal99}  \\
\reac{Mg}{24}{^{4}He}{\gamma}{Si}{28} & \citet{CaughlanFowler_88}  \\
\reac{C}{13}{^{4}He}{n}{O}{16} & \citet{Angulo_etal99}  \\
\reac{O}{17}{^{4}He}{n}{Ne}{20} & \citet{Angulo_etal99}  \\
\reac{O}{18}{^{4}He}{n}{Ne}{21} & \citet{Angulo_etal99}  \\
\reac{Ne}{21}{^{4}He}{n}{Mg}{24} & \citet{Angulo_etal99}  \\
\reac{Ne}{22}{^{4}He}{n}{Mg}{25} & \citet{Angulo_etal99}  \\
\reac{Mg}{25}{^{4}He}{n}{Si}{28} & \citet{Angulo_etal99}  \\
\hline
\end{tabular}
\end{table}

The electron screening factors for all reactions are those from
\citet{Dewitt_etal73} and \citet{Graboske_etal73}.  The abundances of
the various elements are evaluated with the aid of a semi-implicit
extrapolation scheme, as described in \citet{Marigo_etal01}. This
technique, which does not assume nuclear equilibrium, is a convenient
compromise between the higher accuracy typical of the explicit scheme,
and the better stability of the solution guaranteed by the implicit
scheme.

\subsection{Neutrino losses}
\label{sec_neutr}
Energy losses by electron neutrinos are taken from
\citet{Munakata_etal85} and \citet{ItohKohyama_83}, but for plasma
neutrinos, for which we use the fitting formulae provided by
\citet{Haft_etal94}.

\subsection{Convection}

\subsubsection{Overshoot from the convective core}
\label{sec_conv}

The extension of convective boundary of the core is estimated by means
of an algorithm which takes into account overshooting from the central
convective region \citep{Bressan_etal81}.  In our formalism the main
parameter describing the overshoot is the mean free path of convective
bubbles {\em across} the border of the convective region, expressed in
units of pressure scale height, i.e. $\Lambda_{\rm c}$.  Importantly,
this parameter in the \citet{Bressan_etal81} formalism is not
equivalent to others found in literature.  For instance, the
overshooting scale defined by $\Lambda_{\rm c}=0.5$ in the Padova
formalism roughly corresponds to the 0.25 pressure scale height {\em
  above} the convective border, adopted by the Geneva group
\citep[][and references therein]{Meynet_etal94} to describe the same
physical mechanism, i.e.\ $\Lambda^{\rm G}_{\rm c}=0.25$.

It is well known that with this scheme it is not possible to use a
unique choice of the overshoot parameter independently of the stellar
mass. This is particularly true for main sequence (MS) stars in the
transition between models with radiative and convective cores, i.e. in
the mass range between $M\sim1.0\,M_\odot$ and $M\sim1.5M_\odot$,
depending on the adopted chemical composition.  In this mass range, a
strict application of the above overshoot criterion give rise to the
development of a convective core that grows too much with the
evolution, producing a turn-off morphology which is not observed in
the HR diagram \citep[see the discussion in ][]{Aparicio_etal90}.
This is true also for the solar model for which the convective core
would persist up to the present age of 4.6~Gyr, which is not supported
by current interpretation of helioseismology observations.

We thus adopt a variable overshoot parameter $\Lambda_{\rm c}$ to
describe the development of overshoot, in the transition from main
sequence models with radiative and convective cores:
\begin{itemize}
	\item
$\Lambda_{\rm c}$ is zero for stellar masses $M\le M_{\rm O1}$,
\item In the range $M_{\rm O1}\le M\le M_{\rm O2}$, we adopt a gradual
  increase of the overshoot efficiency with mass, from zero to
  $\Lambda_{\rm max}$.  This because the calibration of the
  overshooting efficiency in this mass range is still very uncertain
  due to the scarcity of stellar data provided by the oldest
  intermediate-age and old open clusters.  Some works
  \citep{Aparicio_etal90}, however, indicate that this efficiency
  should be lower than in intermediate-mass stars.
\item For $M>M_{\rm O2}$, we adopt $\Lambda_{\rm c}=\Lambda_{\rm
    max}$.
\end{itemize}

At variance with previous releases \citep{Bressan_etal93,
  Girardi_etal00, Bertelli_etal08}, where the limiting masses $M_{\rm
  O1}$ and $M_{\rm O2}$ were fixed ($M_{\rm O1}=1\,M_\odot$ and
$M_{\rm O2}=1.5\,M_\odot$), we now adopt a transition region that
depends on the initial chemical composition of the model set.  This
allows a better description of the effects of the elemental abundance
on this phenomenon, which is needed because the onset of convective
core overshoot leaves clear signatures that can be tested in
observational Colour-Magnitude diagrams (CMD).  The transition masses
are automatically defined by running a preliminary subset of stellar
tracks without diffusion, to check for the presence of a persistent
convective unstable nucleus, without considering the overshoot (i.e.,
using the Schwarzschild criterium for convective borders).  Since
during the PMS phase the core may be already convective, we define the
lower limit $M_{\rm O1}$ to be the minimum mass that is able to
maintain this unstable core after a $20\,\%$ of the initial hydrogen
fraction has been burned in the center.

\begin{figure}
  \resizebox{\hsize}{!}{\includegraphics{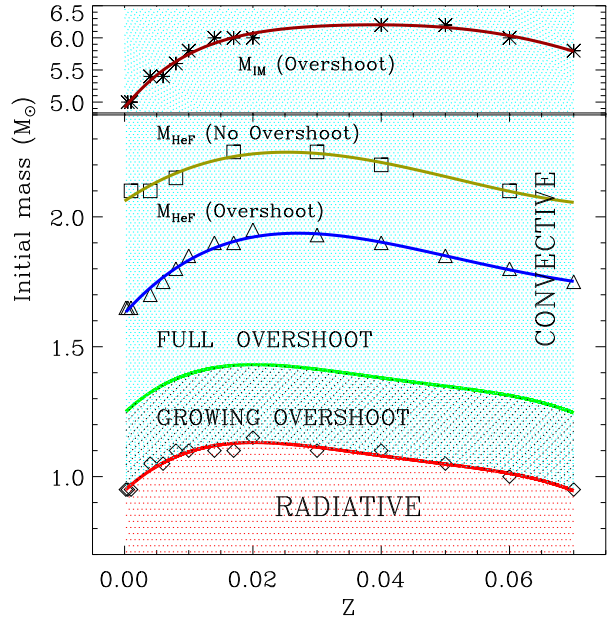}}
  \caption{The behaviour of a few critical masses as a function of
    metallicity, for our scaled-solar models following the
    $Y=0.2485+1.78\,Z$ enrichment law (see Sect.~\ref{sec_chemic}).
    From bottom to top: the minimum mass that maintains a persistent
    convective core during H burning, $M_{\rm O1}$ (see text for more
    details); the mass above which core overshoot is taken at the
    maximum efficiency $M_{\rm O2}$; the minimum mass for a model to
    ignite central He non degenerately, $M_{\rm HeF}$ for both the
    overshoot and the no-overshoot cases; finally, in the upper panel,
    the minimum mass of the stars that ignites C in a non electron
    degenerate core, $M_{\rm IM}$. The typical resolution in
    determining these mass limits is of 0.05~\Msun\ for $M_{\rm O1}$
    and $M_{\rm HeF}$, and of 0.2~\Msun\ for $M_{\rm IM}$. The curves
    are polynomial fits to the corresponding values.}
    \label{fig_masses}
\end{figure}

$M_{\rm O1}$ varies with both metallicity and helium content.  The
bottom panel of Fig.~\ref{fig_masses} illustrates the behaviour of
$M_{\rm O1}$ as a function of metallicity for the main set of models
to be described in this paper (see Sect.~\ref{sec_tracks}).  Notice
that the resolution in the $M_{\rm O1}$ determination is of
0.05~\Msun, which is the typical mass spacing we adopt for low-mass
stars. The minimum values of $M_{\rm O1}$, of $\sim\!0.95$~Msun, occur
at the extremes of very low and very high metallicities.  $M_{\rm O1}$
is found to be $1.1$~\Msun\ at solar metallicities, and reaches a
maximum of 1.15~\Msun\ at $Z=0.02$.  These values are in good
agreement with the lowest mass of a non-diffusive model with a
convective core, equal to 1.14~\Msun, found by \citet{Michaud_etal04}
for assumed solar metallicities of either $Z=0.0175$ or $Z=0.0199$.
The decreasing trend of $M_{\rm O1}$ at lower metallicities can 
be explained by the larger luminosity at given mass, that causes the star to
have a more concentrated nuclear energy source hence a higher radiative gradient
in the core \citep[e.g.][]{Marigo_etal01}.

For the maximum overshooting efficiency we adopt $\Lambda_{\rm
  max}=0.5$, i.e.\ a moderate amount of overshooting, which coincides
with the values adopted in the previous \citet{Bertelli_etal94} and
\citet{Girardi_etal00} models.  This corresponds to about $0.25\,H_P$
of overshoot region {\em above} the convective border found in other
common formalisms.

$M_{\rm O2}$ is always set to be $M_{\rm O1} + 0.3$\,\Msun.  This
choice is motivated by the modelling of the open cluster M\,67 (see
Sect.~\ref{sec_conclu} and Fig.~\ref{fig_m67new}), which indicates
overshooting as efficient as $\Lambda_{\rm c}\simeq0.5$ already at
masses of $\sim\!1.3$~\Msun, for solar-metallicity stars.  This choice
is also supported by the SMC cluster NGC~419
\citep[Sect.~\ref{sec_conclu} and ][]{Girardi_etal09, Kamath_etal10},
in which the turn-off probes masses between $\sim\!1.65$ and
1.9~\Msun.

Furthermore, we assume that the overshooting efficiency $\Lambda_{\rm
  c}$ increases linearly with mass between $M_{\rm O1}$ and $M_{\rm
  O2}$, in order to ensure a smooth enough transition between the
properties of stars with radiative and convective cores. This
prescription has to be considered as a conservative approach, that
will be revisited in future works (e.g.\ Rubele et al., in prep., and
Rosenfield et al., in prep.).  The theoretical difficulties in
defining the efficiency of overshooting in this transition region are
well-known \citep[e.g.][]{Aparicio_etal90}.  From the observational
side, the recent indications from asteroseismology are still
ambiguous: while the observations of $\alpha$\,Cen~A
\citep{deMeulenaer_etal10} suggest negligible overshooting in
solar-metallicity stars of mass $\sim\!1.1$~\Msun, recent
asteroseismology studies of the nearby old low-mass star HD~203608
\citep{Deheuvels_etal10}, with $[Z/X]\simeq-0.5$, indicate the action
of overshooting (with $\alpha_{\rm ov}=0.17$, which corresponds to
$\alpha_{\rm ov}\simeq0.32$ in our formalism) at masses as low as
$0.95$~\Msun, which is probably just slightly above the $M_{\rm O1}$
limit. Clearly, the behaviour of overshooting in the transition region
from $M_{\rm O1}$ to $M_{\rm O2}$ deserves more detailed
investigations.

Another important difference with respect to previous releases of
Padova evolutionary tracks is that the region of overshoot is
considered to be radiative.  In our original scheme this region was
assumed to be adiabatic while the majority of other authors took the
overshoot region to be radiative.  With the former assumption the
model adjusts the internal radiative temperature gradient in such a
way that the unstable region is smaller than in the case of a
radiative overshoot region.  Thus the effects of this new assumption
are an overall larger mixed core.  From a numerical point of view this
assumption provides a faster convergence, because the original
formalism causes some instability during the convergence process, that
must be controlled.

In the stages of core Helium burning (CHeB), the value $\Lambda_{\rm
  c}=0.5$ is used for all stellar masses. This amount of overshooting
dramatically reduces the extent of the breathing pulses of convection
found in the late phases of CHeB \citep[see][]{Chiosi_etal92}.

\subsubsection{Overshoot in the convective envelope}
\label{sec_conv2}

\citet{Alongi_etal91} were the first to consider the possibility that
the base of the convective envelope can give rise to a sizable
overshoot region.  In the past, two important observational effects
have been studied in relation with this phenomenon, namely the
location of the RGB Bump in the red giant branch of low mass stars
(globular clusters and old open clusters) and the extension of the
blue loops of intermediate mass stars.  Both effects have been found
to be better explained by a moderate amount of overshoot, with a
typical extension {\em below} the border of about $0.25-1.0\,H_P$.
Against this possibility it has always been argued that the
calibration of the solar model do not require a sizable overshoot
region because the transition between the fully adiabatic envelope and
the radiative underlying region in our Sun, is already well reproduced
by models without overshoot.  This however does not exclude the
possibility that just below the fully adiabatic region convection may
penetrate in form of radiative fingers that are even able to induce a
significant mixing.  Very recently it has been argued that a mechanism
of this kind could even provide a better agreement with the physical
state of matter in this transition region derived from solar
oscillations data \citep{Christensen-Dalsgaard_etal11}.  The size of
this effect has been recently estimated to be $\Lambda_{\rm
  e}\sim0.4\,H_P$ but it is also consistent with a larger value
$\Lambda_{\rm e}\sim0.6\,H_P$ which is in very good agreement with the
one adopted since \citet{Alongi_etal91}.  As shown in
Sect.~\ref{sec_sun} below , our own solar calibration suggests a value
of $\Lambda_{\rm e}(Adiabatic)\le0.05\,H_P$.  We have not yet included
the criterion adopted by \citet{Christensen-Dalsgaard_etal11}.  In
view of the above results we will maintain in the present paper the
the same prescriptions adopted by \citet{Alongi_etal91} and
\citet{Bressan_etal93}: $\Lambda_{\rm e}=0.05$ in the envelope of
stars with $M<M_{\rm O1}$ and $\Lambda_{\rm e}=0.7$ in the envelope of
stars with $M>M_{\rm O2}$.  In between we adopt a smooth transition
similar to that used for the central overshoot in between these two
limiting masses.  Moreover, the temperature gradient in the envelope
overshoot region is kept equal to the radiative one.

\begin{figure}
  \resizebox{\hsize}{!}{\includegraphics{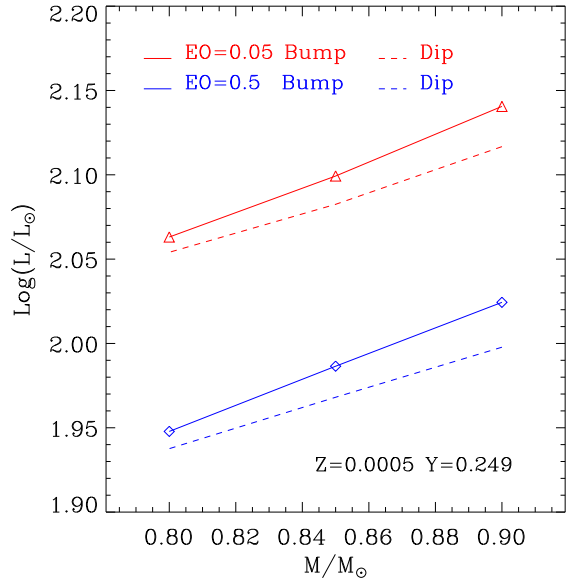}}
  \caption{The luminosity both at the maximum (solid lines) and
    minimum (``dip'', dashed lines) of the bump feature along the
    first-ascent RGB, for a few low-mass tracks of
    $(Z=0.0005,Y=0.249)$ computed with $\Lambda_{\rm e}=0.05\,H_P$
    (red) and $\Lambda_{\rm e}=0.5\,H_P$ (blue).  In models with high
    $\Lambda_{\rm e}$ the RGB bump is typically $\sim\!0.3$~mag
    fainter than in model with negligible envelope overshooting.}
    \label{fig_rgbbump}
\end{figure}

It is worth recalling that, contrary to the overshoot in the central
regions, the envelope overshoot has almost no effects on the
evolutionary properties of the stars, though it may affect some
observable properties, as already discussed.  In this respect, an
important observational evidence that we will investigate in the next
future is the effect of the envelope overshoot on the abundance of
light elements (e.g. $^7$Li), during and after the PMS evolution.  To
this purpose one should also take into account that early main
sequence stars could still suffer from a non negligible mass accretion
\citep{DeMarchi_etal11} and that envelope overshoot, during the PMS
phase, could be more efficient than in the main sequence phase
\citep{Molaro_etal12}. However this is beyond the scope of this paper
and will be addressed in a forthcoming investigation.

Another important effect of envelope overshooting is that of lowering
the luminosity level of the bump along the first-ascent evolution of
low-mass stars.  Fig.~\ref{fig_rgbbump} compares the bump luminosity
between models computed with $\Lambda_{\rm e}=0.05\,H_P$ (red) and
$\Lambda_{\rm e}=0.5\,H_P$ (blue) for a few low-mass tracks of
composition $(Z=0.0005,Y=0.249)$.  In models with the higher
$\Lambda_{\rm e}$, the RGB bump is typically $\sim\!0.3$~mag fainter
than in models with negligible envelope overshooting. By comparison,
the observed RGB bump in globular clusters is about 0.2 to 0.4~mag
fainter than that predicted by models without envelope overshooting
(\citealt{DiCecco_etal10}; see also \citealt{Cassisi_etal11}), though
at the higher metallicities this result depends on the adopted
metallicity scale. This simple comparison favours larger values of
$\Lambda_{\rm e}$ in low-mass metal-poor stars, than here adopted
based on the solar calibration.  Needless to say, this argument will
be the subject of further calibrations.

\subsubsection{Breathing Pulses and Semiconvection}
During the end of the core-He burning phase, models undergo a well
known instability, the breathing pulses.  In a series of pulses of the
helium core a non negligible fraction of the helium in the surrounding
stable regions is brought and burned into the core. This causes an
increase of the helium burning lifetime which is not very significant
by itself. But this has a dramatic effect on the duration of the
subsequent AGB phase, which can be shortened by a significant factor.
Given that this bears on the luminosity function of the brightest M
and C stars, it is important to deal with this mechanism in the more
accurate and physically sound way as possible.  At present there is
no way to resolve the issue of whether this effect is numerical or
real, because it happens independently from the adopted mixing scheme,
i.e. with mild overshoot or with semi-convection.  Work is in progress
to calibrate the efficiency of this mechanism with observations.

\subsubsection{Temperature gradient in the convective region}
The energy transport in the convective regions is described according
to the mixing-length theory of \citet{mlt}.  The super-adiabatic
gradient is maintained until its difference with respect to the
adiabatic one decreases below $\nabla_{\rm Element}-\nabla_{\rm Adi} <
10^{-6}$. While in the convective core this condition is always
fulfilled, in the convective envelope it is fulfilled only near the
bottom of the region. The mixing length parameter $\alpha_{\rm MLT}$
is fixed by means of the solar model calibration performed below (see
\refsec{sec_sun}), and turns out to be $\alpha_{\rm MLT}=1.74$.  As
already stated the temperature gradient in any overshoot region is
assumed to be the radiative one.

\subsection{Mass meshing and other }
The stellar structure equations are solved maintaining the algorithm
of the fitting method.  The fitting mass is typically set at $M/M_{\rm
  Tot}=1-10^{-5}$, but it can be regulated on the basis of the
corresponding temperature or depending on other needs.  An adaptive
mesh is used in the inner structure with a number of points that is
about 1500 during hydrogen burning, 3000 during He burning and 5000
during the thermally-pulsing AGB (TP-AGB).

As for the atmosphere we adopt a plane-parallel grey model where the
temperature stratification is given by a modified Eddington
approximation for radiative transport:
\begin{equation}
T^{4} = \frac{3}{4} T_{\rm eff}^{4} \left [  \tau +q\left ( \tau  \right ) \right ]
\label{eq_ttau}
\end{equation}
where $\tau(r)$ is the optical depth, and $q(\tau)$ is the Hopf
function.  Assuming the gas pressure $P$ as the independent variable,
the differential equation $d \tau /d \log P $ is solved by using a
standard predictor-corrector method, starting from $\tau=0$, where the
gas pressure vanishes and the total pressure is concides with the
radiation pressure, to $\tau=\tilde{\tau}$, the latter being the
optical depth of the photosphere.  The overall method is fully
described in \citet{Kipp_etal67}.

We explicitly follow the evolution of 26 elements, listed in
Sect.~\ref{sec_rates}.  The network of nuclear reactions is solved
with a semi-implicit extrapolation scheme \citep{BaderDeuflhard83}
that treats all nuclear reactions together and without any assumption
on nuclear equilibria \citep[see][]{Marigo_etal01}. At each time step
the abundance equations are solved after each Henyey convergence, and
particular care is paid both to mantain sufficiently small abundance
variations (by adopting a suitably short time step), and to preserve
the number of nucleons.

\subsection{Diffusion}

Microscopic diffusion is included following the implementation by
\citet{Salasnich99}.  The diffusion coefficients are
calculated following \citet{Thoul_etal94} and the corresponding system
of second order differential equations is solved together with the
chemistry equation network, at the end of each equilibrium model.
Diffusion is applied to all the elements considered in the code in the
approximations that they are all fully ionized.

As already extensively discussed by \citet{Chaboyer_etal01}, our
knowledge of the diffusion process is still incomplete because if
applied with the same diffusion constants used for the Sun in metal
poor stars of similar mass, they lead to surface abundance that are in
disagreement with observations.  Indeed, on the basis of accurate
abundance determinations in stars of globular cluster NGC~6397
\citep{Gratton_etal01}, \citet{Chaboyer_etal01} concluded that
microscopic diffusion should be fully inhibited in the external layers
of metal poor stars at least down to a depth of about 0.005~\Msun\
from the photosphere and partially inhibited down to 0.01~\Msun\ from
the photosphere.

In the case of metal rich stars, microscopic diffusion in the external
layers of stars of this masses is already inhibited by the more
extended external convection.  In general it is believed that, when
one considers more massive stars with less extended surface convective
regions, the effects of diffusion are negligible because of the much
shorter stellar evolutionary times.  Even in this case, however, it
can be shown that if the mesh spacing in the external layers is kept
suitably small for accuracy purposes, diffusion can noticeably change
the surface composition even for masses well above those of globular
clusters if the external convection disappears.  This in turn changes
the interior opacities and produces somewhat unrealistic paths in the
HR diagram, indicating that diffusion must be inhibited at the surface
of these stars.

Other criteria can be used to deal with the particular problem of
inhibiting molecular diffusion in those circumstances where its
application would lead to evident discrepancies with observations.
Surface abundances in stars without extended surface convective
regions are very sensitive to the effects of atomic diffusion and
radiative accelerations \citep{Turcotte_etal98}.  However in general
it is found that some sort of extra-mixing beyond the base of the
external convective layers must be added in order to reproduce the
observed abundances \citep[e.g.][]{Richer_etal00}.  This extra-mixing,
of unknown origin but whose effect is that of moderating/inhibiting
other diffusive processes, is parameterized as a turbulent diffusion
with a coefficient that is calibrated on the observed surface
abundances of old stars \citep[see e.g.][ and references
therein]{Vandenberg_etal12}.  Unfortunately, when the calibrating
observable is the surface Li abundance, the results of different
investigations, based on different stars, do not agree
\citep{Melendez_etal10,Nordlander_etal12}.  Furthermore, this
calibration is challenged even more by the discovery that early main
sequence stars still suffer from a non negligible mass accretion
\citep{DeMarchi_etal11}. This tail of accretion, preceded by a very
efficient turbulent mixing during the pre-main sequence phase that
completely destroy Li, could reshape our view of the surface evolution
of this element \citep{Molaro_etal12}, and likely require another
different calibration of the inhibiting mechanism.  In view of all the
above reasons and since we neglect for the moment radiative levitation
\citep{Vauclair83}, we switch off the diffusion process when the mass
size of the external convection falls below a threshold mass that we
assume to be $\Delta{M_{\rm conv}}=$0.5\,\% of the total stellar mass.
Moreover diffusion is not considered in stars that develop a
persistent convective core, i.e.\ when core overshoot is taken into
account.

\begin{figure*}
  \resizebox{0.4\hsize}{!}{\includegraphics{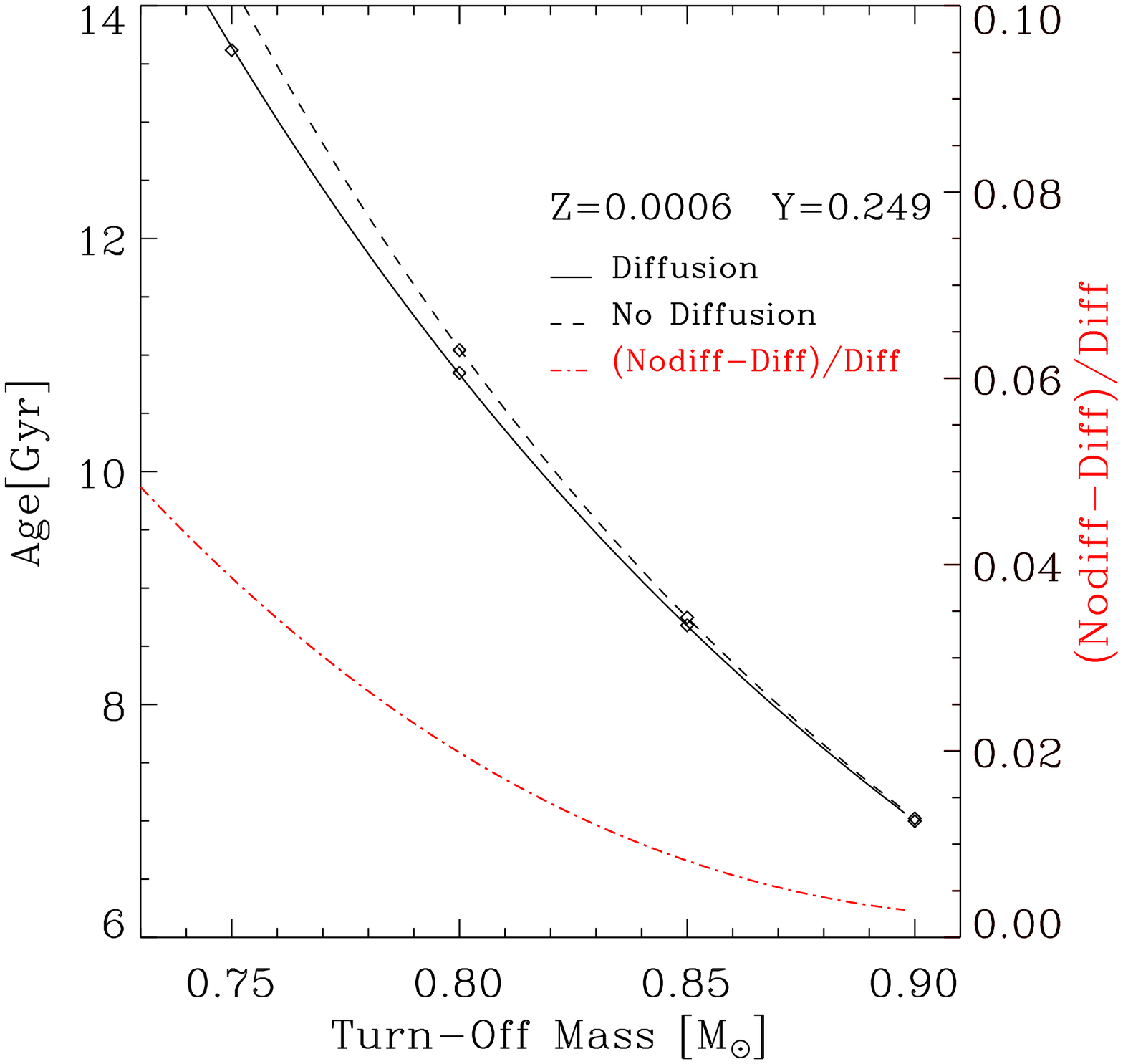}}
  \resizebox{0.4\hsize}{!}{\includegraphics{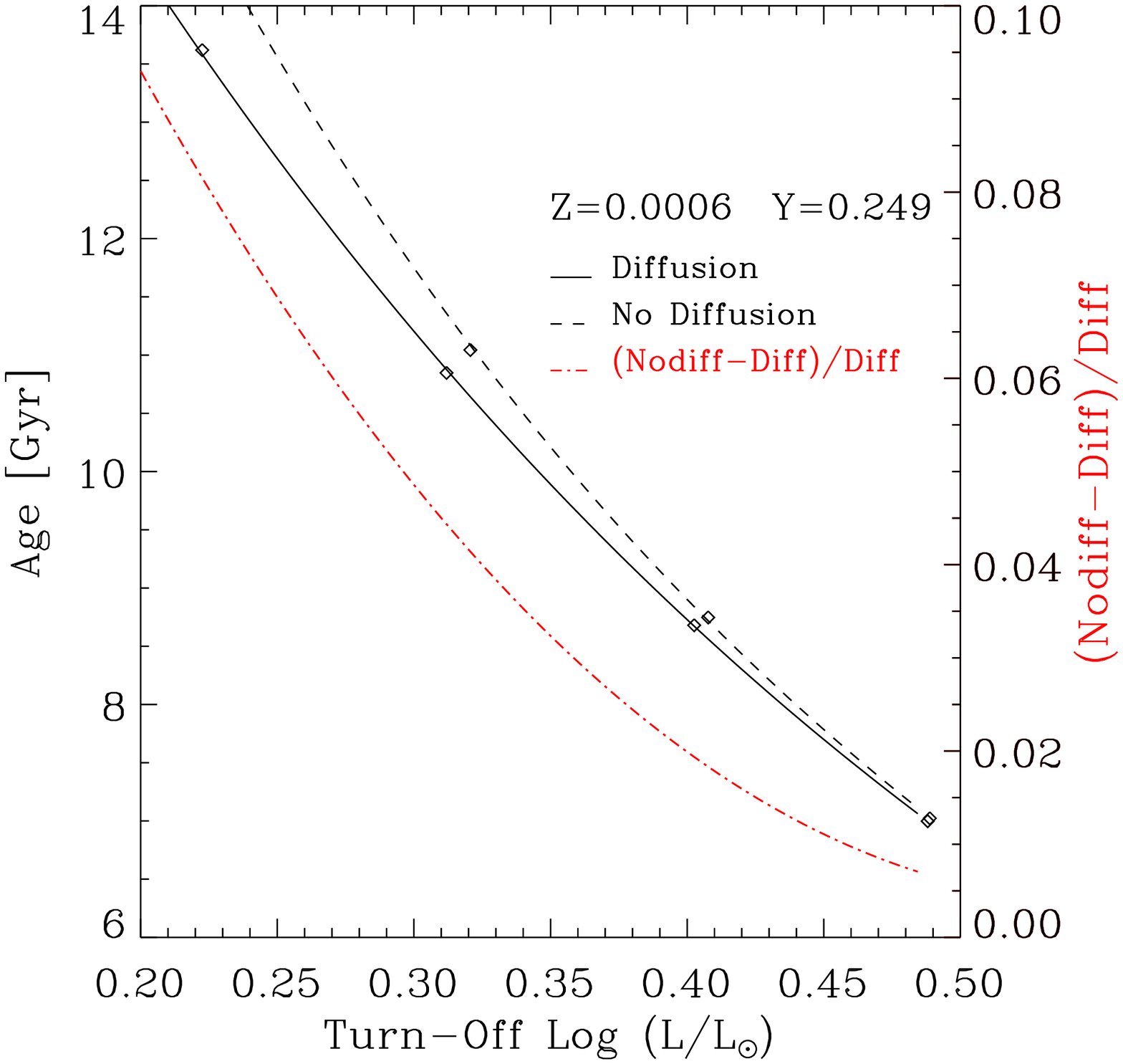}}
  \caption{Comparison between the turn-off age as a
    function of the stellar mass for models with and without diffusion (dark
    lines); with the relative difference plotted in red (see scale at
    the right axis). The right panel shows the same as a function of
    the turn-off luminosity.  Differences between diffusive and
    non-diffusive models become negligibly small (less than 0.5~\% in
    age) for masses above $\sim\!0.9$~\Msun.}
    \label{fig_diffusioneffect2}
\end{figure*}

\begin{figure}
  \resizebox{\hsize}{!}{\includegraphics{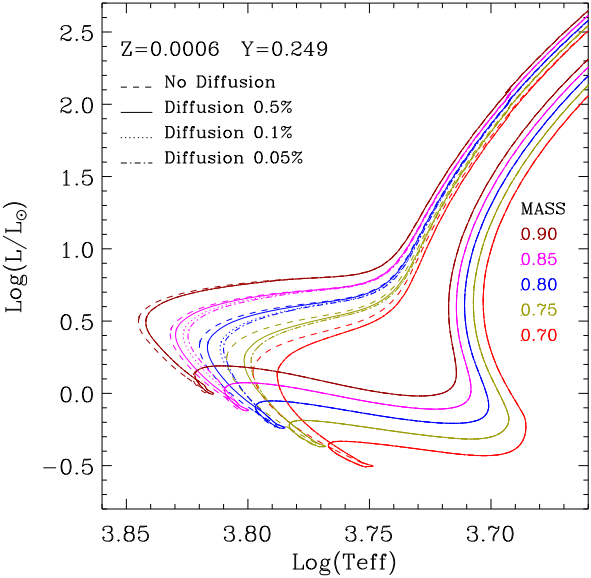}}
  \caption{Comparison in the HR diagram between evolutionary tracks
    computed with different thresholds for turning-off diffusion, for a
    set of tracks of low masses and for a low metallicity of
    $(Z=0.0006, Y=0.249)$. Diffusion is switched off as soon as the
    mass size of the surface convection falls below the indicated fractional values of the total mass.
    For reference, the  case without diffusion is also shown. See text for more details.
  }
  \label{fig_diffusioneffect1}
\end{figure}

Figure \ref{fig_diffusioneffect2}
illustrates the impact of diffusion on the hydrogen burning lifetimes
in low-mass metal poor evolutionary tracks with chemical composition
$(Z=0.0006, Y=0.249)$.
From this figure we see that, with this criterion,  the effect of
diffusion becomes very small for masses above $\sim\!0.9$~\Msun, which
corresponds to ages of about 6~Gyr.
At the 12~Gyr ages typical of globular
clusters (with turn-off masses slightly below $\sim\!0.8$~\Msun), the
differences between diffusive and non-diffusive models become
significant: for instance, for the same turn-off luminosity, ages
derived from non-diffusive models would be about 8~\% older than those
derived from diffusive models. These differences are perfectly in line
with those found by other authors \citep[][and references
therein]{CastellaniScilla99}.

To explore the sensitiveness of the results to different values of the
threshold for turning-off diffusion, we have computed also the cases
with $\Delta{M_{\rm conv}}=$0.1\,\% and 0.05\,\% for the same chemical
composition $(Z=0.0006, Y=0.249)$.  All the above cases are displayed
in Fig. \ref{fig_diffusioneffect1}.  Decreasing the threshold mass
delays the quenching of diffusion and produces a larger shift of the
turnoff region toward cooler temperatures.  This effect is more
pronounced in the mass range between $0.85$~\Msun\ and $0.75$~\Msun.
At lower masses surface convection is always well developed while at
higher masses convection rapidly disappears.  However, we notice that
in the latter case this result is also due to the inhibition of
diffusion throughout the whole star, an assumption that could be
improved following recent investigations that underline the importance
of turbulent mixing in the more external regions \citep[see
e.g.][]{Vandenberg_etal12}.  In the light of these considerations we
plan a revision of the tracks in this mass and metallicity range in
the near future.

\section{Calibration of the solar model}
\label{sec_sun}

The comparison with the solar model is a necessary step to check the
quality of the input physics and to calibrate some parameters that
cannot be derived from the theory.

\begin{table}
\caption{Solar calibration.}
\begin{tabular}{l l l r}
\hline\hline
              Solar data      & Value    & error & source\\
\hline
$L_\odot$ (10$^{33}$erg\,s$^{-1}$) &   3.846     & 0.005       & \citet{Guentheretal1992}   \\
$R_\odot$ (10$^{10}$\,cm)        &   6.9598    & 0.001    &  \citet{Guentheretal1992} \\
$T_{{\rm eff}_,\odot}$ (K)       &   5778   & 8    &   from $L_\odot$ \& $R_\odot$\\
$Z_\odot$                        &   0.01524   & 0.0015    & \citet{Caffau_etal11} \\
$Y_\odot$                        &   0.2485    & 0.0035   &  \citet{BasuAntia04}\\
$(Z/X)_\odot$                    &   0.0207    & 0.0015    &  from $Z_\odot$ \& $Y_\odot$ \\
$R_{\rm ADI}/R_\odot$            &   0.713     & 0.001    &  \citet{BasuAntia97}\\
$\rho_{\rm ADI}$                 &   0.1921      & 0.0001    &  \citet{Basuetal2009}\\
$C_{\rm S, ADI}/10^7$cm/s        &   2.2356      & 0.0001     & \citet{Basuetal2009}\\
\hline\hline
Model & Tab-EOS & Fly-EOS& \\
\hline
Age$^+$(Gyr)           &4.593      & 4.622   &  \\
$Z_{\rm initial}$      &0.01774    & 0.01774$^*$    &  \\
$Y_{\rm initial}$      &0.28       & 0.28$^*$    &  \\
$\alpha_{\rm MLT}$     &1.74       & 1.74$^*$    &  \\
$\Lambda_{\rm e}$     &0.05        & 0.05$^*$    &  \\
\hline
$L$ (10$^{33}$erg\,s$^{-1}$)    &   3.848  & 3.841    &  \\
$R$ (10$^{10}$\,cm)   &  6.9584   & 6.96112   &  \\
$T_{{\rm eff}}$ (K)   &  5779   & 5775    &  \\
$Z_{\rm S}$            &0.01597    & 0.01595    & \\
$Y_{\rm S}$            &0.24787    & 0.24762   &\\
$(Z/X)_\odot$          &0.02169    & 0.02166   &\\
$R_{\rm ADI}/R_\odot$ &0.7125      & 0.7129    &\\
$\rho_{\rm ADI}$      &0.1887      & 0.1881    &\\
$C_{\rm S, ADI}/10^7$cm/s &2.2359      & 2.2364     &\\
\hline
\end{tabular}
\label{table:scalib}
\par Distribution of heavy elements from \citet{Caffau_etal11}.
$^+$ Age includes the pre main sequence phase.
$^*$Values taken from the calibration obtained with tabulated EOS.

\end{table}
\begin{figure*}
\begin{center}
\includegraphics[width=0.45\textwidth]{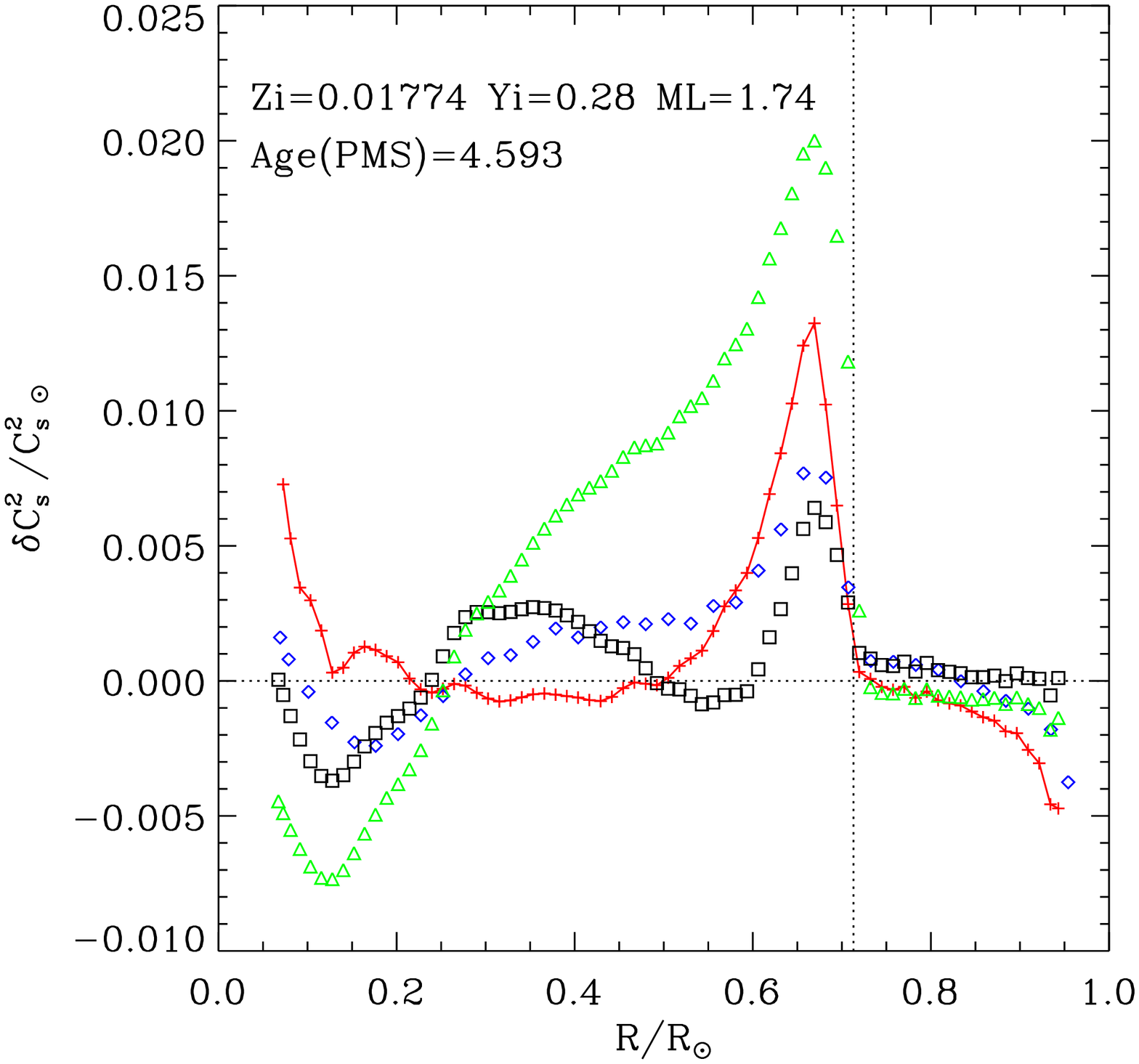}
\includegraphics[width=0.45\textwidth]{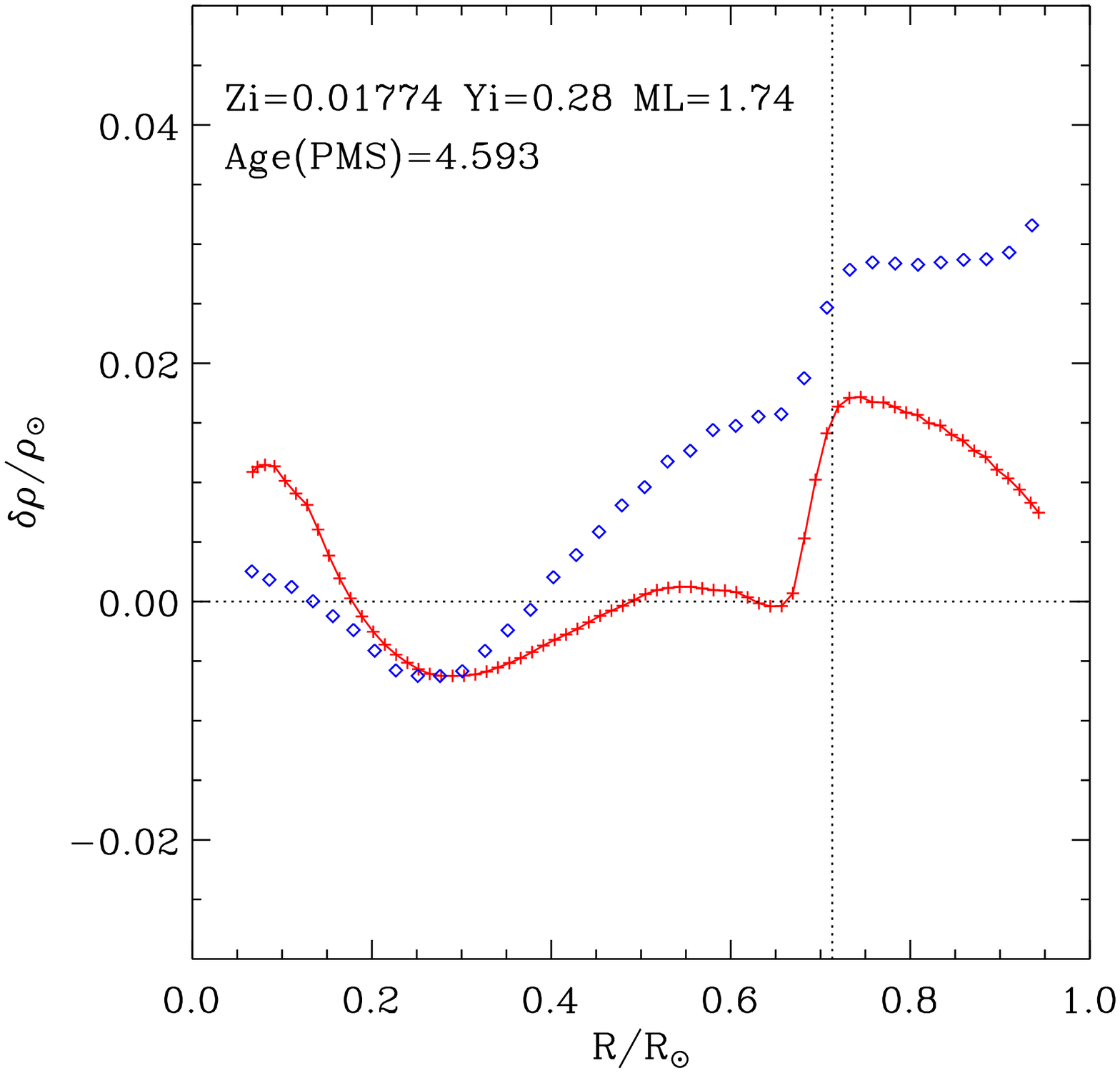}\\
\hfill
\caption{Solar model obtained using the tabulated Free-EOS.  Left
  panel: interior relative differences (in the sense Sun minus model)
  in the squared sound speed $\delta c_{\rm s}^2/c_{\rm s}^2$.  Solar
  values are obtained from MDI data \citep{Basu_etal00}.  Symbols are
  as follows.  Red crosses: our model; blue diamonds:
  \citet{Basuetal2009} model differences with respect to MDI data
  using \citet{GrevesseSauval_98} solar abundances; black squares:
  \citet{Serenelli_etal09} using \citet{GrevesseSauval_98} solar
  abundances; green triangles: \citet{Serenelli_etal09} using
  \citet{Asplund_etal09} solar abundances.  Right panel: interior
  relative differences for the density profile.  In this case only the
  \citet{Basuetal2009} model is shown for comparison.  $Z_i$, $Y_i$
  and ML refer to the initial metallicity and helium abundance and to
  the MLT parameter of our model, respectively.  The quoted age
  includes the PMS lifetime which is about 40~Myr, determined by the
  stage on the ZAMS where the total gravitational luminosity falls,
  and remains, below a few per cent.  Values obtained from this
  calibration are shown in Table \ref{table:scalib}.}
\label{fig_calib_eos_tab}
\end{center}
\end{figure*}

For this purpose we have computed a large grid of tracks of one solar
mass, from the PMS phase to an age of 4.8~Gyr, and varying the initial
composition of the Sun, $Z_{\rm initial}$ and $Y_{\rm initial}$, the
mixing length parameter $\alpha_{\rm MLT}$ and the extent of the
adiabatic overshoot at the base of the convective envelope
$\Lambda_{\rm e}$.  To calibrate these free parameters, the models
were compared with a set of solar data obtained from the literature
which are summarized in Table~\ref{table:scalib}.  The reference solar
data used here come from MDI observations described in
\citet{Basu_etal00}.  More recent data can be found in
\citet{Basuetal2009} from BiSON experiment.  However since we wish to
compare our model with previous ones in literature, in particular
considering the effects of changing the present day surface solar
composition, we have found more convenient to use \citet{Basu_etal00}
data, with respect to which there are several comparison at different
chemical composition made in literature.  Besides verifying the
performance of our solar model another goal of the comparison is to
obtain the mixing length parameter, $\alpha_{\rm MLT}$, that will be
used to compute all the stellar evolutionary sets.  For this reason
the calibration has been obtained exactly with the same set-up used
for the calculations of the other tracks, i.e. with tabulated EOS and
opacities and, of course, using microscopic diffusion.  However in
order to check the effects of interpolating the EOS we have also
computed a solar model with the ``on-the-fly'' version of FreeEOS.  In
this model we have adopted the parameters of the best fit obtained
with the tabulated EOS, and we changed only the solar age in order to
match as much as possible the solar data. The parameters of our best
model are listed in the lower part of Table~\ref{table:scalib}.

In the left panel of Fig.~\ref{fig_calib_eos_tab} we plot the relative
variation of the squared sound speed $\delta c_{\rm s}^2/c_{\rm s}^2$
as a function of the fractional radius inside the Sun (red crosses and
solid line).  The sense of the comparison is (Sun$-$model)/Sun in both
panels.  The Solar values were obtained from MDI data
\citep{Basu_etal00}.  In the same figure we also show: the
\citet{Basuetal2009} model differences with respect to MDI data,
obtained with the \citet{GrevesseSauval_98} solar abundances (blue
diamonds); the \citet{Serenelli_etal09} difference profile using
\citet{GrevesseSauval_98} solar abundances (black squares) and using
\citet{Asplund_etal09} solar abundances (green triangles).  In the
right panel we show the relative variation of the density profile (red
crosses and solid line) and we compare it only with the
\citet{Basuetal2009} difference profile (blue diamonds).  Notice that
the quoted age includes the PMS lifetime, which when determined by the
stage on the ZAMS where the total gravitational luminosity goes to
zero, amounts to about 40 Myr.  Inspection of Table~\ref{table:scalib}
and of Fig.~\ref{fig_calib_eos_tab} shows that our solar model
performs fairly well, taking also into account that lowering the solar
metallicity from the \citet{GrevesseSauval_98} values, had a severe
impact in the performance of the solar model.  Particularly
interesting are the small values of the $\delta c_{\rm s}^2/c_{\rm
  s}^2$ and density profiles in the central radiative regions of the
Sun, indicating that good agreement is possible also with abundances
lower than those of \citet{GrevesseSauval_98}.  Toward the central
region our model suffer of a a slightly lower sound speed and a
slightly lower density than shown by the solar data. A similar though
less pronounced problem is also present in the comparison of models
with Solar data extracted from the Bison experiment
\citet{Basuetal2009}.

For the track computed with the ``on-the-fly'' FreeEOS we did not
repeat the calibration process thus it does not reproduce the basic
solar data of Table~\ref{table:scalib} as well as in the previous
case.  However we see that we may also obtain a fairly good agreement
with solar data, by slightly increasing the age.

From the initial values of the metallicity and Helium abundance of the
Sun, $Y_{\rm initial}, Z_{\rm initial}$, and adopting for the
primordial He abundance $Y_{\rm p }=0.2485$
\citep{Komatsu_et_al_2011}, we obtain also the helium-to-metals
enrichment ratio, $\Delta Y/\Delta Z=1.78$.  This value will be used
in the following to determine the default value of helium content for
any given global metallicity.

\section{Initial chemical composition of the evolutionary tracks}
\label{sec_chemic}

As already specified in \refsec{distribution} our selected reference
distribution of heavy elements is taken from \citet{GrevesseSauval_98},
except for a subset of species for which we adopt the recommended
values according to the latest revision by \citet{Caffau_etal11}.

For the present release, other distributions have been considered such as
those typical of the Large Magellanic Cloud, of massive early-type
galaxies and of $\alpha$-depleted elements.

The helium and metal mass fractions are chosen according to the
relation $Y=0.2485+1.78\,Z$ obtained from the solar calibration. Sets
with varying helium at fixed metallicity are also considered.

We remind here that when we change the fractional abundance by number
of heavy elements to obtain a new distribution, their fractional
abundance by mass is re-normalized in such a way that the global
metallicity, $Z$, is kept constant.

Full sets of opacity and EOS tables are recomputed for each different
chemical composition. Chemical compositions for which we have already
computed full sets of evolutionary tracks are shown in
Table~\ref{table:chemzy}. More sets are being computed for
$\alpha$-enhanced distributions.
\begin{table}
  \caption{Initial chemical composition of the sets ready at the time of this writing}
  \label{table:chemzy}
  \begin{tabular}{l l l | c c c}
    \hline\hline
    &   &       & \multicolumn{3}{c}{$[\alpha/{\rm Fe}]$ (dex)} \\
    \cline{4-6}
    Z      & Y & [M/H]$^{\rm a}$ & $0$ & $+0.2$ & $+0.4$ \\
    \hline
    0.0005& 0.2485 & $-1.49$ & $\times$ & $\times$ &          \\
    0.001 &  0.250 & $-1.19$ & $\times$ & $\times$ &          \\
    0.004 &  0.256 & $-0.58$ & $\times$ & $\times$ &          \\
    0.005 &  0.258 & $-0.48$ &          &          & $\times$ \\
    0.006 &  0.259 & $-0.40$ & $\times$ & $\times$ &          \\
    0.008 &  0.263 & $-0.28$ & $\times$ & $\times$ &          \\
    0.010 &  0.267 & $-0.18$ & $\times$ & $\times$ & $\times$ \\
    0.010 &  0.263 & $-0.18$ &          & $\times$ &          \\
    0.014 &  0.273 & $-0.02$ & $\times$ &          &          \\
    0.015 &  0.276 & $-0.01$ &          & $\times$ &          \\
    0.017 &  0.279 & $+0.06$ & $\times$ & $\times$ &          \\
    0.017 &  0.350 & $+0.11$ & $\times$ &          &          \\
    0.017 &  0.400 & $+0.15$ & $\times$ &          &          \\
    0.02  &  0.284 & $+0.14$ & $\times$ & $\times$ & $\times$ \\
    0.03  &  0.302 & $+0.34$ & $\times$ & $\times$ &          \\
    0.04  &  0.321 & $+0.48$ & $\times$ & $\times$ &          \\
    0.05  &  0.339 & $+0.60$ & $\times$ & $\times$ &          \\
    0.06  &  0.356 & $+0.70$ & $\times$ & $\times$ & $\times$ \\
    0.07  &  0.375 & $+0.78$ & $\times$ & $\times$ &          \\
    \hline
  \end{tabular}
  \vspace{2pt}\\ $^{\rm a}$Approximated value from $[{\rm M/H}]=\log((Z/X)/0.0207)$.
\end{table}

\begin{figure*}
  \resizebox{1.03\textwidth}{!}{\includegraphics{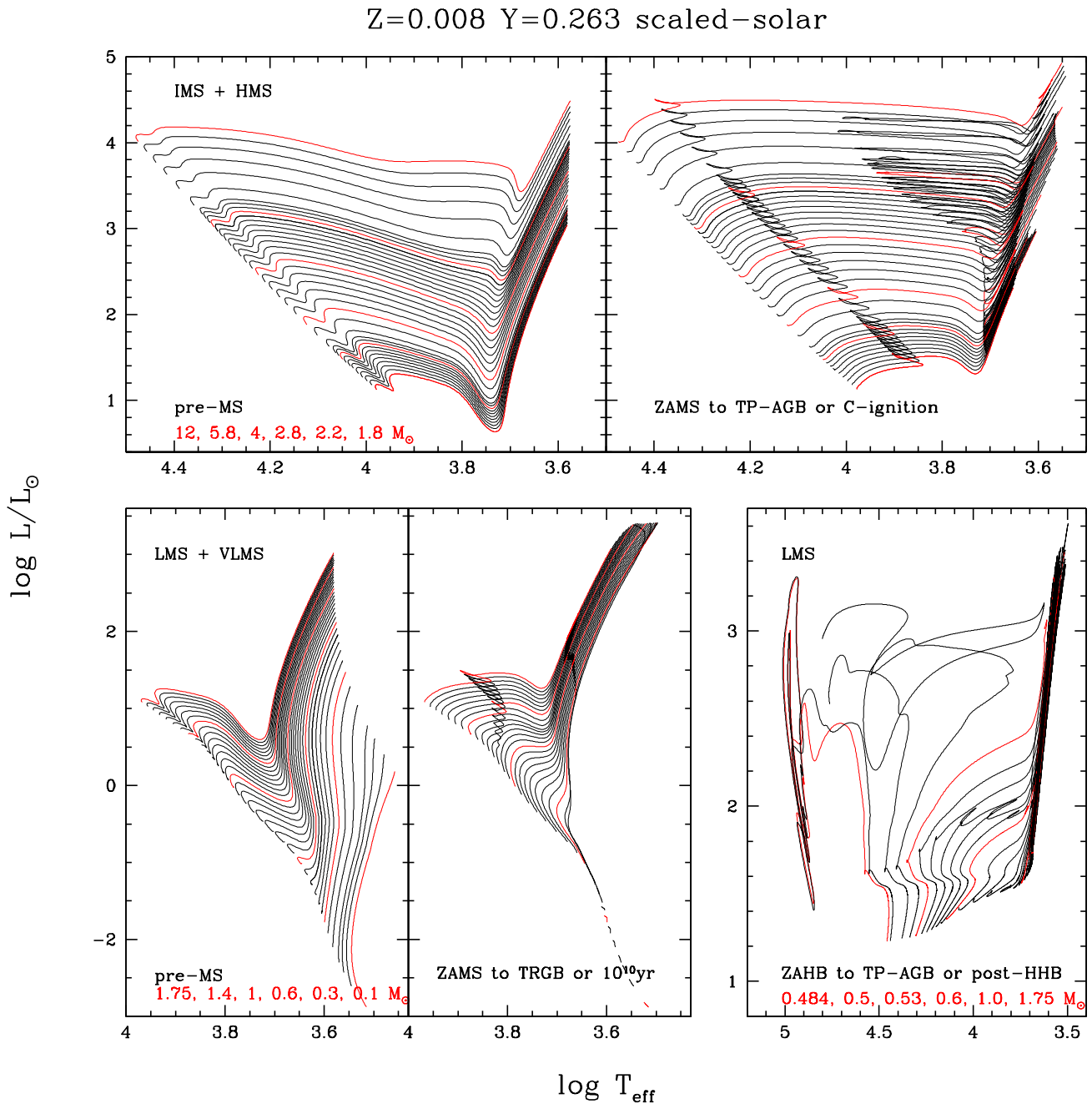}}
  \caption{Evolutionary tracks in the HR diagram for the selected
    chemical composition $Z=0.008, Y=0.263$ and with $[\alpha/{\rm
      Fe}]=0$. {\bf The upper panels} show the tracks for
    intermediate- and high-mass stars (IMS+HMS) split into the PMS and
    later evolution; in the latter case the evolution is followed
    until the beginning of the TP-AGB, or up to C-ignition for
    $M>5.8$~\Msun. The low and very-low mass stars (LMS+VLMS) are
    shown in the{\bf bottom-left panels}. In this case the evolution
    after the PMS is plotted either up to the tip of the RGB, or up to
    an age of 100~Gyr for $M<0.6$~\Msun. The {\bf bottom-right panel}
    shows the complete suite of low-mass He-burning tracks computed
    for this set. The evolution starts at the ZAHB sequence and goes
    up to either the beginning of the TP-AGB for $M>0.6$~\Msun, or or
    up to the stages of post-early-AGB, and AGB-manqu\'e (and in some
    cases including the initial dimming towards the white dwarf
    sequence), typical of lower-mass HB stars that follow a hot-HB
    evolution (HHB).}
\label{HR}
\end{figure*}

\section{Stellar tracks}
\label{sec_tracks}

A complete set of evolutionary tracks in the HR diagram is shown in
Fig.~\ref{HR}, for the chemical composition $Z=0.008, Y=0.263$ with
$[\alpha/{\rm Fe}]=0$. Such plots are available for all sets listed in
Table~\ref{table:chemzy}.

\subsection{Mass range}
In this paper we present the evolution of stars with initial mass
between 0.1~\Msun\ and 12~\Msun.  Higher stellar masses will be
considered in a future paper.  The mass spacing is smaller than in
previous releases, being typically of 0.05~\Msun\ in the range of low
mass stars and $\sim\!0.2$~\Msun\ for intermediate-mass stars.  This
allows a good description of the transition between low- and
intermediate-mass stars, and between intermediate- and high-mass
stars.

\subsection{Pre Main Sequence}
The PMS evolution is a new characteristic of this release and it is
implemented as follows.  A stellar model with suitable mass and
initial composition is obtained along the Hayashi track and
artificially expanded until its central temperature falls below
$T_{\rm c} \approx 10^5$~K.  At this stage no nuclear reactions are
active and the model is fully convective and homogeneously mixed.
From this configuration, we allow the star to evolve at constant mass
along the PMS contraction and subsequent phases.  The PMS phase of
$Z=0.008, Y=0.263,[\alpha/{\rm Fe}]=0$ tracks is shown in
Fig.~\ref{HR}.

\begin{figure}
  \resizebox{\hsize}{!}{\includegraphics{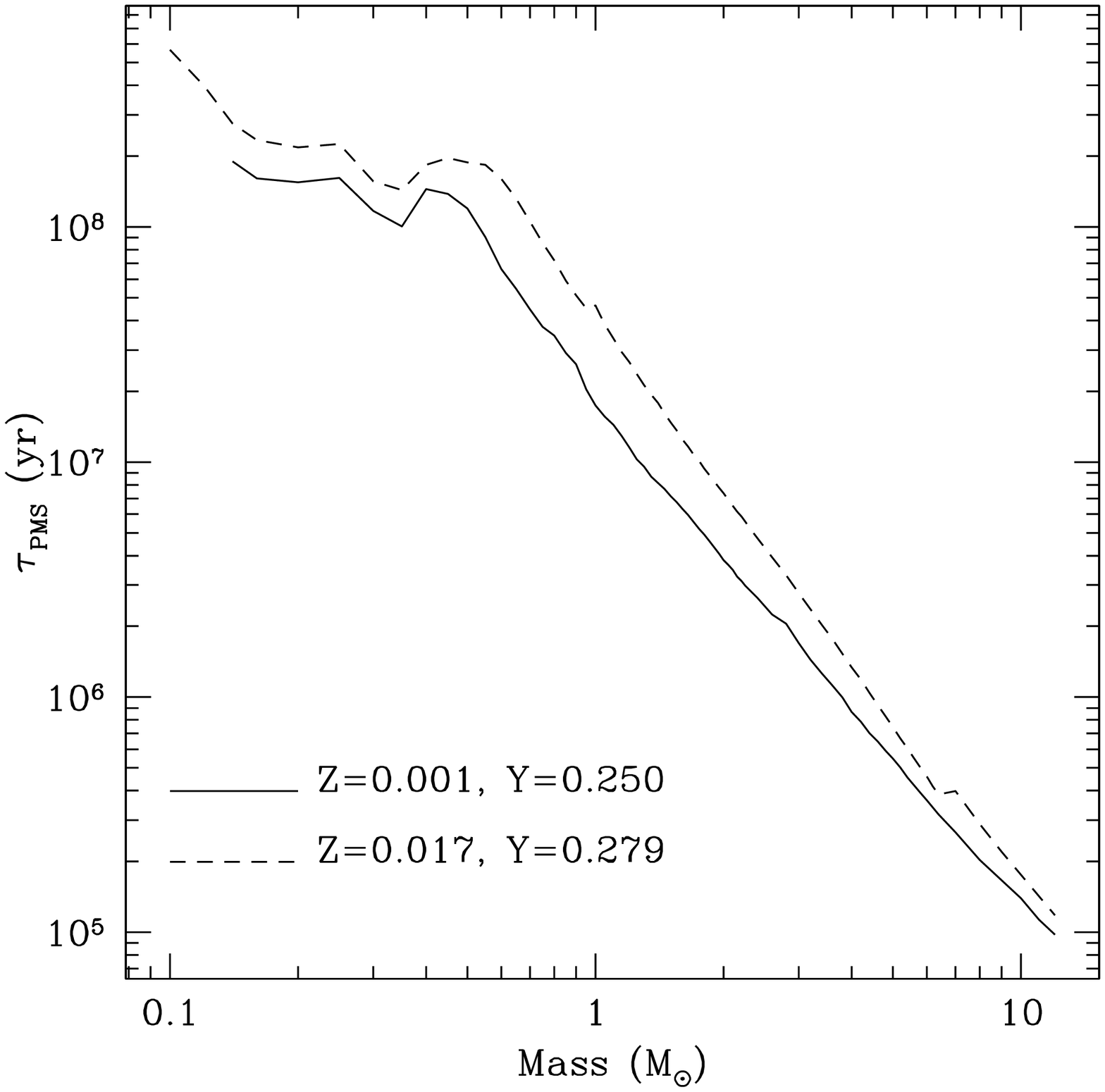}}
  \caption{PMS lifetimes, as a function of stellar mass and for two
    different metallicities.}
  \label{fig_pms}
\end{figure}

The typical PMS lifetimes are illustrated in Fig.~\ref{fig_pms}, as a
function of stellar mass and for two different metallicities. It can
be noticed that the behaviour as a function of mass is very regular
just for masses higher than about 1~\Msun. For smaller masses, the
higher relative importance of the deuterium burning, and the onset of
completely convective stars at $M\la0.4$~\Msun, cause the behaviour to
be less monotonic.  Also, it should be mentioned that the definition
of what is PMS phase is not so straightforward, so that in some cases
we find some small glitches in the mass--PMS-lifetime relation, even
at higher masses.  These features seem to be less of a problem while
making the isochrones for the PMS phase.

\subsection{Other evolutionary stages}
\label{sec_massranges}

During the PMS evolution a number of low temperature reactions change
the abundance of some light elements and the model is no longer
homogeneous.  The end of this phase is also characterized by a rapid
reassessment of the internal structure during which the evolutionary
track in the HR diagram shows characteristic patterns (loops) that
depend on the initial mass and on the initial chemical composition.
The above effects render difficult the identification of the Zero Age
Main Sequence (ZAMS) point in an unambiguous way. To define the
beginning of the MS we thus consider the point where the
evolutionary speed in the HR diagram shows an abrupt drop, decreasing
by more than two order of magnitudes in a very short time.  This point
is clearly detectable in all tracks and it will also be used to define
the beginning of the MS when constructing the corresponding
isochrones.

Very low mass stars with mass $M\leq0.45$~\Msun\ are evolved up to an
age of 100~Gyr. Some test models have been evolved up to the central
Hydrogen exhaustion and subsequent core degenerate phase, when the
model abandon the Red Giant Branch (RGB) because the Hydrogen shell
reaches the surface of the star. These tracks show some interesting
features but of course are not of any interest for single stellar
evolution because the age where these features happens are by far
larger than the Hubble time.

Low mass stars, with mass between $M\sim0.5$~\Msun\ and an upper value
($M_{\rm HeF}$) which mainly depend on the chemical composition and on
the eventual extra mixing in the convective nucleus, ignite Helium in
an electron degenerate hydrogen-exhausted core and undergo the so
called He-flash. When the He-burning luminosity exceeds the stellar
surface luminosity by a factor of 200, the track is interrupted and
the following evolution along the central Helium burning phase is
computed separately. The He burning phase of low mass stars is
re-started from a suitable Zero Age Horizontal Branch (ZAHB) model
with the same core mass and internal and surface chemical composition
as the last RGB model. The evolution is thus followed up to well
developed thermal pulses on the AGB.  The subsequent TP-AGB evolution
up to the complete ejection of the envelope will be presented in an
accompanying paper (Marigo et al., in preparation).

\begin{figure}
  \resizebox{\hsize}{!}{\includegraphics{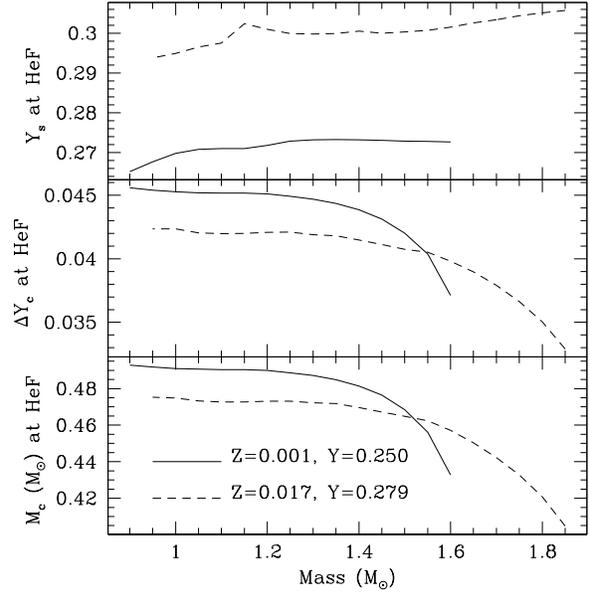}}
  \caption{Quantities that primarily determine the properties of
    He-burning low-mass stars, as a function of mass and for two
    solar-scaled chemical compositions. {\bf Bottom panel:} The core
    mass at the moment of the He-flash.  {\bf Middle panel:} The same
    for the mass fraction of helium, $\Delta Y_{\rm c}$, that is
    burned into carbon to provide the energy to lift electron
    degeneracy in the core, between the RGB tip and ZAHB models.  {\bf
      Top panel:} The same for the mass fraction of helium in the
    envelope.}
  \label{fig_hef}
\end{figure}

In the construction of the initial ZAHB model we take into account
that a fraction of the helium in the convective core has been burned
into carbon during the He-flash. This fraction is estimated as the
amount of nuclear fuel necessary to lift the core degeneracy, and is
computed for each track. It closely follows the total mass of the core
at the HeF, as illustrated in Fig.~\ref{fig_hef}.

For masses larger than $M_{\rm HeF}$, we follow the evolution either
until the first few (5--20) well developed thermal pulses along the
AGB (in which case the star belongs to the range of intermediate-mass
stars), or until central carbon ignition for higher mass models (in
which case the star is broadly defined a massive star). For a given
input physics (i.e. opacities, neutrino losses, etc.) the separation
mass between the two classes, $M_{\rm IM}$, depends on the chemical
composition and degree of convective core extra mixing.

The values of $M_{\rm HeF}$ and $M_{\rm IM}$ as a function of
metallicity are illustrated in Fig.~\ref{fig_masses} for our grids of
scaled-solar tracks. As can be noticed, the behaviour of $M_{\rm HeF}$
closely follows the one of $M_{\rm O1}$ (Sect.~\ref{sec_conv}), with a
maximum at intermediate metallicities. About the same happens for
$M_{\rm IM}$, but with the maximum values -- of $\sim\!6$~\Msun\ --
moving to higher metallicities.

\subsection{$\alpha$-enhanced mixtures}

\begin{figure*}
\begin{minipage}{0.46\textwidth} 
\resizebox{\hsize}{!}{\includegraphics{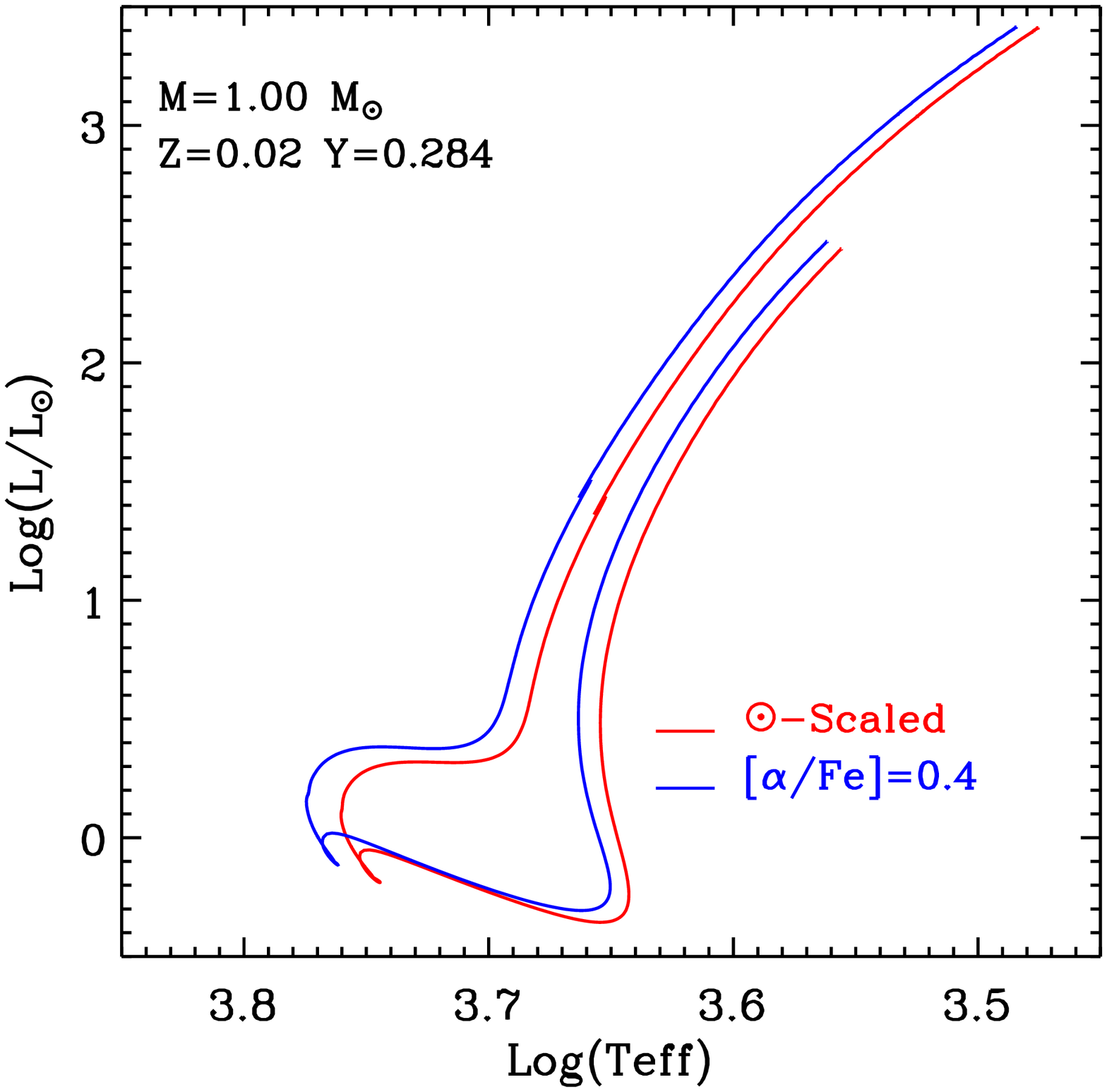}}
\end{minipage}
\hfill
\begin{minipage}{0.46\textwidth} 
\resizebox{\hsize}{!}{\includegraphics{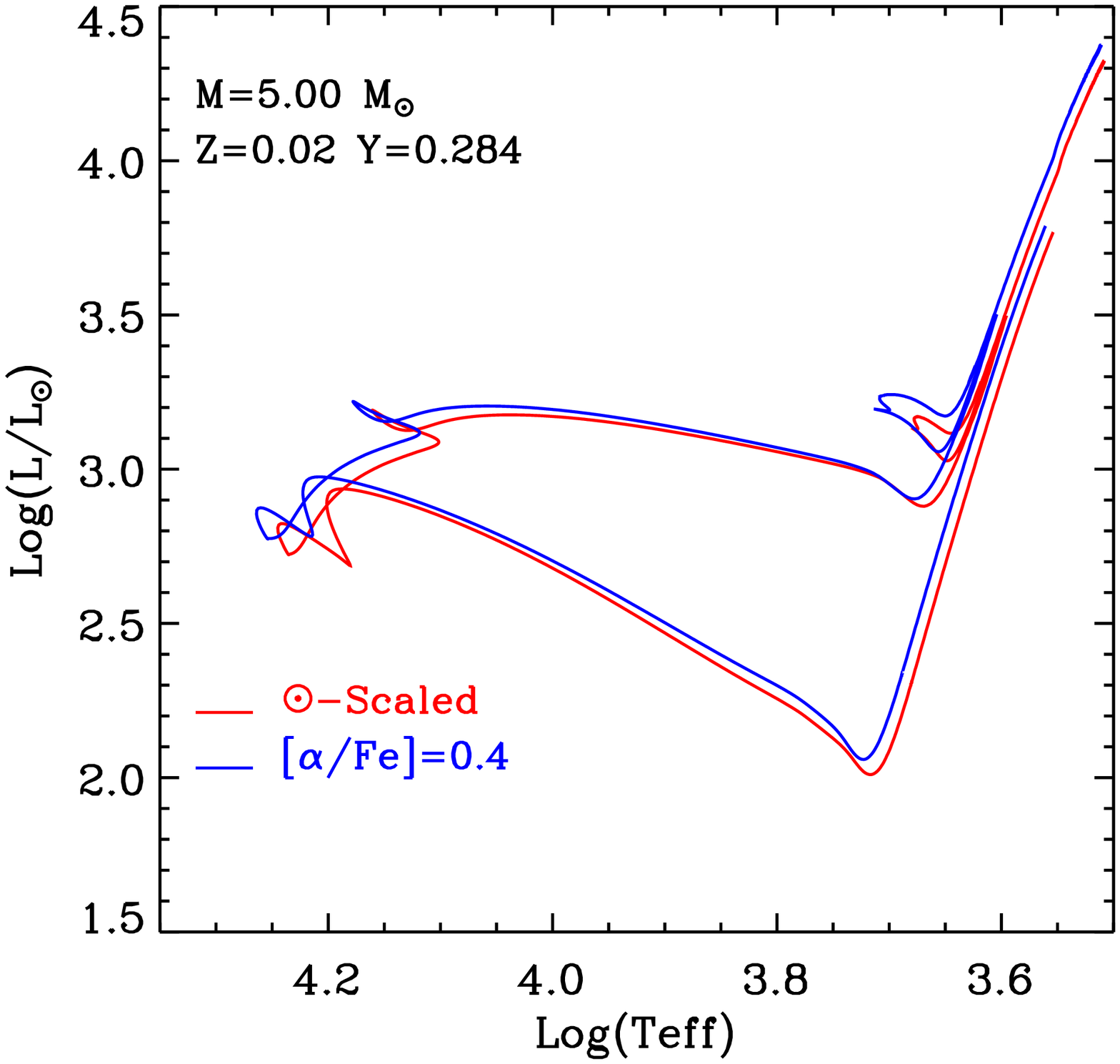}}
\end{minipage}
\caption{Comparison between evolutionary tracks computed either with
  scaled-solar chemical composition , or $\alpha$-enhanced mixtures,
  while keeping the same initial total metallicity Z and helium
  content $Y$. Stellar masses (in $M_{\odot}$) are indicated in each
  plot.}
\label{fig_enhtracks}
\end{figure*}

Figure~\ref{fig_enhtracks} exemplifies the effect of adopting
$\alpha$-enhanced mixtures on the evolutionary tracks in the H-R
diagram.  There is a systematic trend of the $\alpha$-enhanced tracks
(blue line) to be somewhat warmer than the corresponding scaled-solar
cases (red line), especially along the RGB.

In this respect, we call attention to the fact that the effect of the
$\alpha$-enhancement depends critically on the way the chemical
mixture is effectively built.  In general, given an $[\alpha/{\rm
  Fe}]$ ratio, one has two options: a) either keeping the ratio [Fe/H]
fixed while increasing the absolute abundances of the $\alpha$
elements, which leads to a net increase of the total metallicity $Z$;
or b) keeping the metallicity $Z$ fixed while depressing the
abundances of the Fe-group elements and somewhat enhancing those of
the $\alpha$ elements.  The tracks presented in
Fig.~\ref{fig_enhtracks} are computed with the b) option.  In the
latter case the most relevant consequence is the effective increase of
O, Mg, Ne, etc., while in the former case the most important effect is
the depletion of the Fe-group elements.

The changes in the abundances may produce important effects on
evolutionary tracks, mainly due to opacity effects.  We recommend to
refer to \cite{MarigoAringer_09} for a thorough discussion of this
issue (in particular their sections 4.3, 4.4 and figures 20, 21, 22).

As a general rule we may expect that $\alpha$-enhanced tracks computed
according to the a) option tend to be cooler than the corresponding
scaled-solar tracks \citep[see e.g.][]{Vandenberg_etal12} because of a
net increase in the metallicity, while $\alpha$-enhanced tracks
computed according to the b) option tend to be cooler than the
corresponding scaled-solar tracks because the depression of the
Fe-group elements implies a reduction of the H$^{-}$ opacity,
important absorption source at temperatures $\approx 3000-6000$ K. In
fact the H$^{-}$ opacity is highly sensitive to the number of free
electrons, a significant fraction of which is just provided by
Fe-group elements.

\begin{figure*}
\begin{minipage}{0.3\textwidth} 
\resizebox{\hsize}{!}{\includegraphics{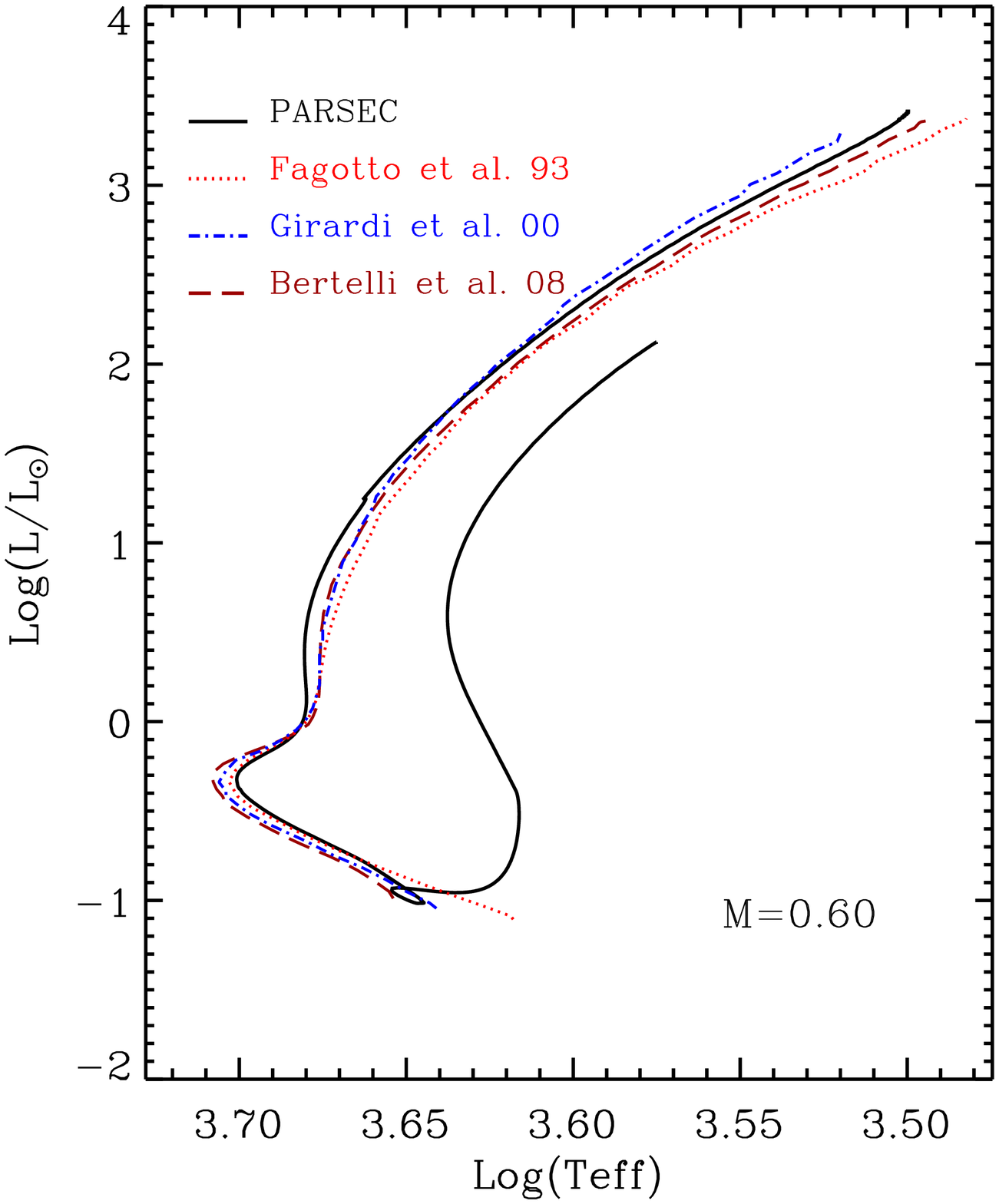}}
\end{minipage}
\hfill
\begin{minipage}{0.3\textwidth} 
\resizebox{\hsize}{!}{\includegraphics{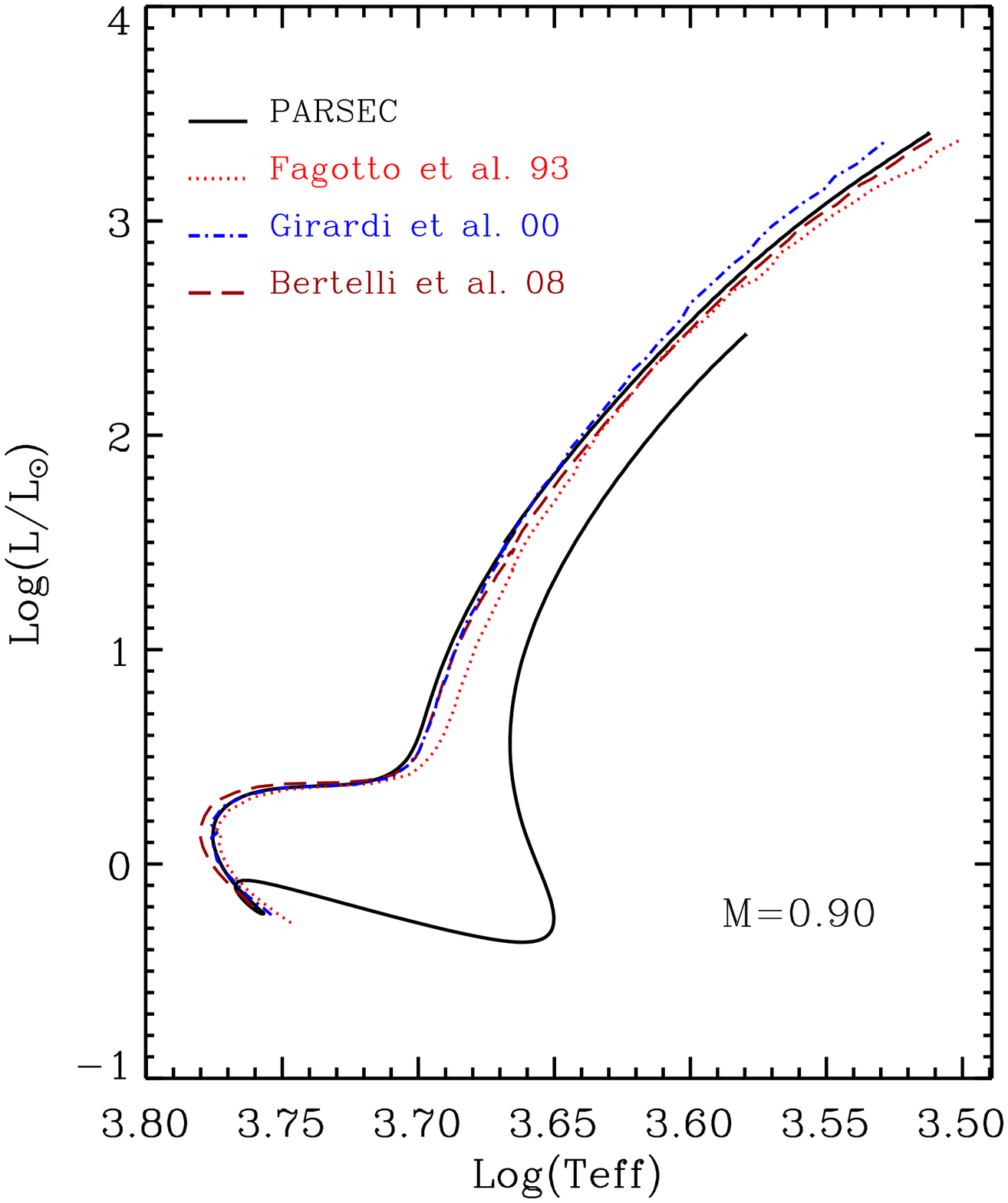}}
\end{minipage}
%
\hfill
\begin{minipage}{0.3\textwidth} 
\resizebox{\hsize}{!}{\includegraphics{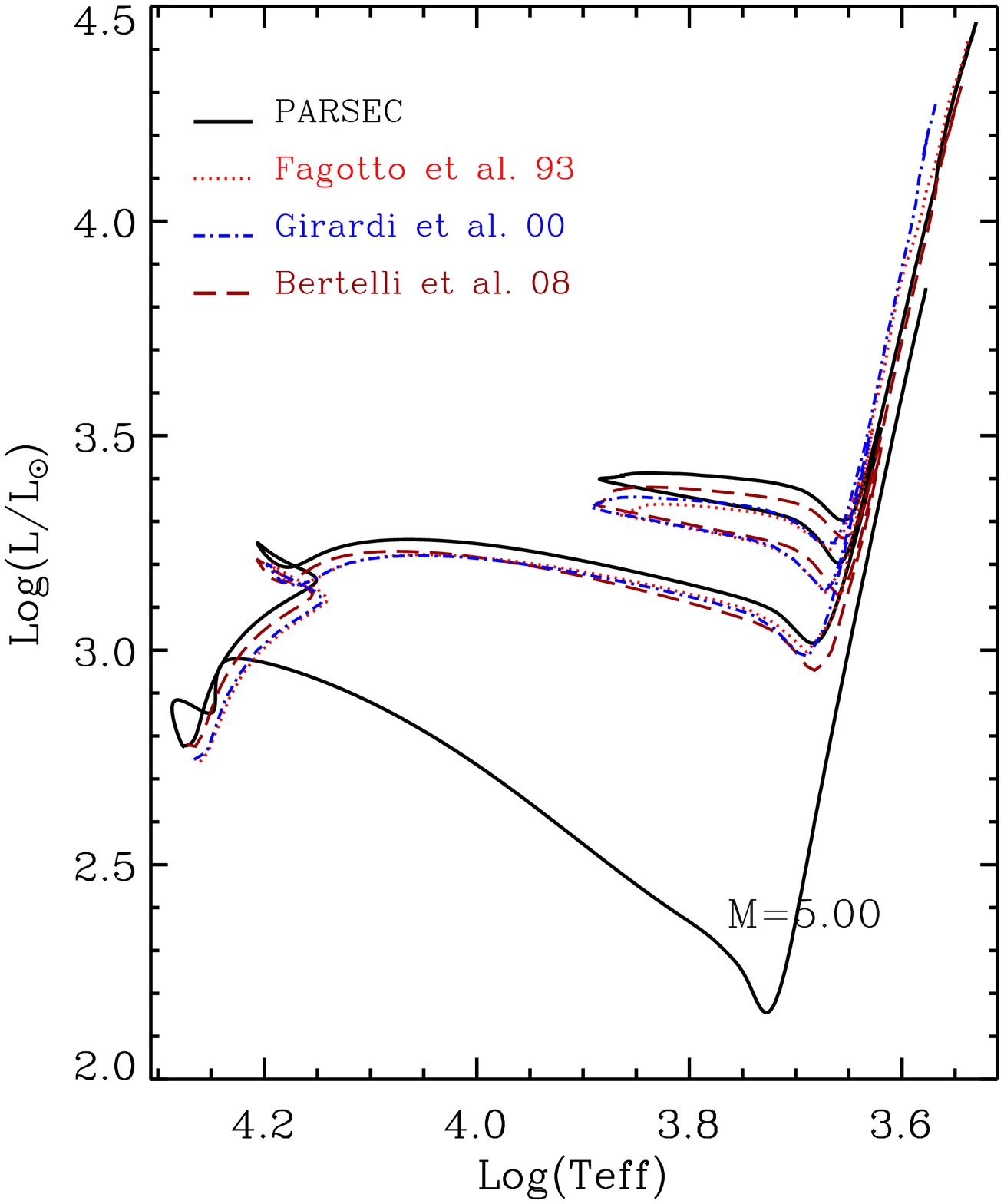}}
\end{minipage}
\vskip 0.2truecm
\caption{Comparison with previous releases of Padova evolutionary
  tracks with $Z=0.008$. Notice that He abundances are slightly
  different among the different sets illustrated: $Y=0.263$ for
  PARSEC; $Y=0.25$ for \citet{Fagotto_etal94}; $Y=0.25$ for
  \citet{Girardi_etal00}; $Y=0.26$ for \citet{Bertelli_etal08}.
}
\label{fig_comparison}
\end{figure*}

\subsection{Comparison with previous releases}
\label{sec_comparison}

In Fig.~\ref{fig_comparison} we compare our new tracks with the
corresponding ones selected from a few previous releases from Padova,
in order to give the reader an immediate impression of the
similarities and differences brought by the updates in input physics.
The selected composition is that typical of young populations in the
Large Magellanic Cloud. The metallicity is $Z=0.008$ for all the
tracks but the Helium abundance $Y$ may be slightly different among
the different sets.

Beside the presence of the PMS in the new tracks, a few general trends
can be seen in the figure.

Fist of all, the effect of diffusion is clearly visible in the tracks
of lower mass.  If we allow for a small shift in the ZAMS caused by
the different input physics, we see that the new track runs cooler
during the MS, and have a cooler turn-off.  Another evident and
important difference in low mass stars is that the base of the RGB is
bluer in the new tracks.  The two effects above will clearly affect
the age determination of old populations, whenever the effective
temperatures are used to constrain their ages.

Moreover, with respect to \citet{Girardi_etal00}, the RGB itself is
less steep and runs at lower temperatures.  In this phase the
differences with respect to \citet{Bertelli_etal08} are less marked
and the tracks have a very similar slope.

In the intermediate mass stars ($M=5$~\Msun) the models show
significant differences already from the MS.  These differences are
mainly due to the larger He abundance of the new tracks, and we limit
the comparison only to the \citet{Bertelli_etal08} set, which has
almost the same composition.  In this case we can see that the new
track runs slightly more luminous not only in the MS but also in the
post-MS phases.  This is mainly due to the different assumption about
the value of the real temperature gradient in the overshoot region.
\citet{Bertelli_etal08} assume that the overshoot region is adiabatic.
The consequence of this assumption in the equilibrium structure is
that the Schwarzschild unstable core is less extended and thus it
renders the extra mixing (with the same overshoot parameter) less
efficient.  In other words assuming that the overshoot region is
radiative is equivalent to assuming a slightly larger overshoot
parameter and, correspondingly, the star evolves at a higher
luminosity because of the larger mixed core.

We notice again that the red giant phase -- in this case the early AGB
-- in the new tracks is significantly cooler than in
\citet{Girardi_etal00}.  This difference is important because it bears
on the mass-loss rates and thus on the subsequent evolution of the
stars along the TP-AGB.

\begin{figure*}
  \resizebox{\hsize}{!}{\includegraphics{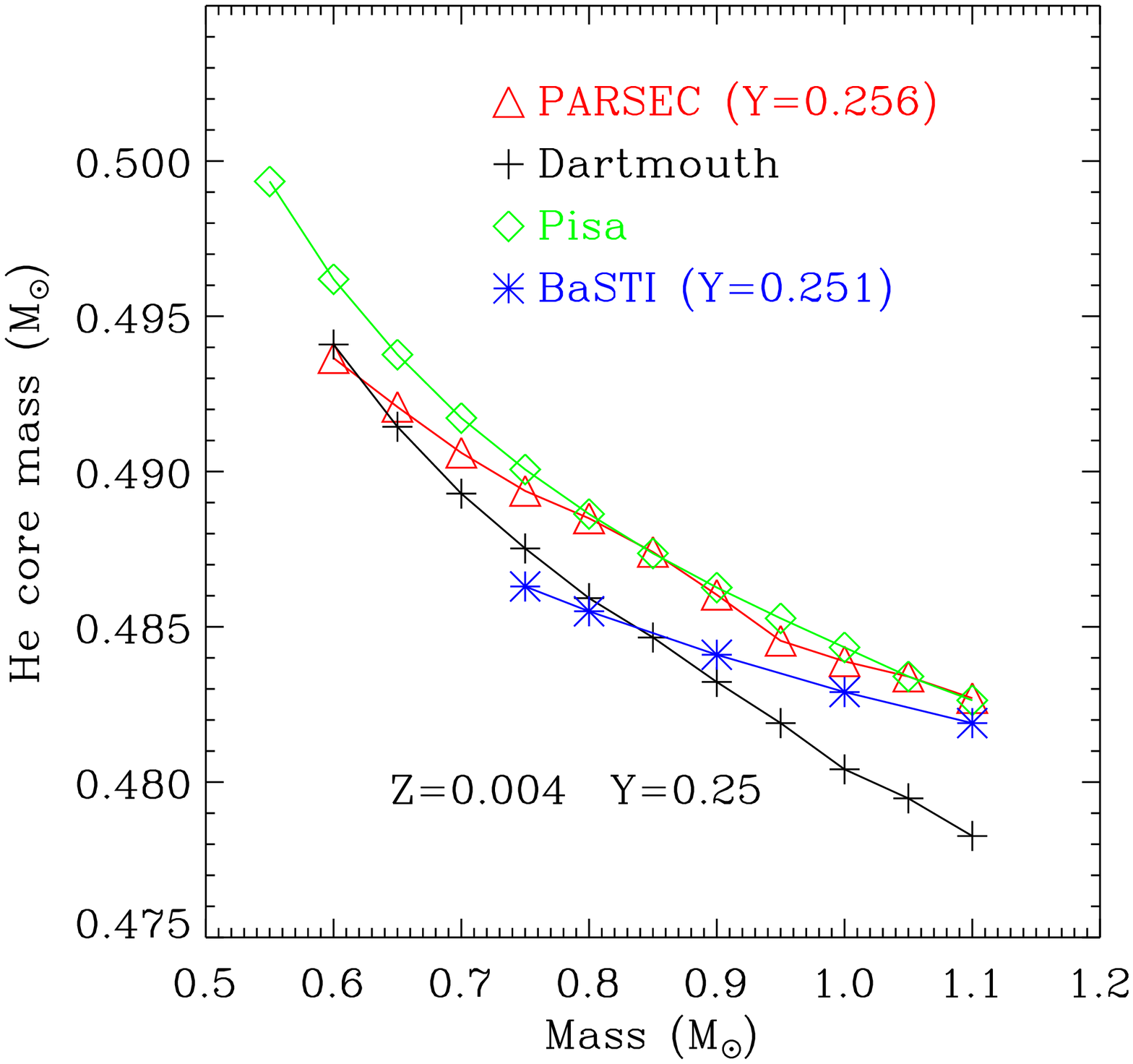}
  \includegraphics{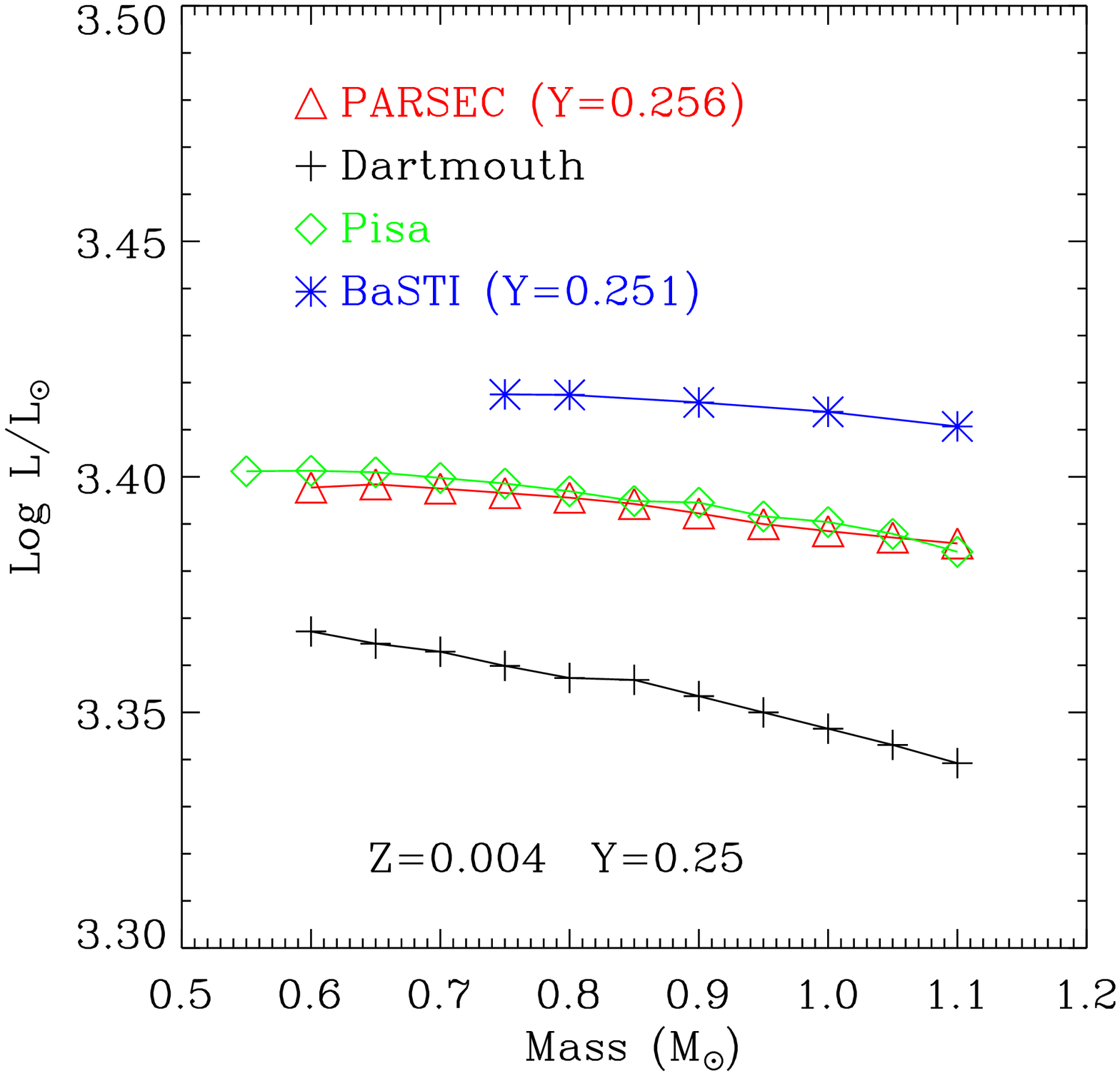}}
\caption{He core mass (left panel) and luminosity (right panel) at the
  tip of the red giant branch. The PARSEC models for $Z=0.004,
  Y=0.256$ are compared with those from other authors, namely
  \citet[][ Dartmouth]{dartmouth08}, \citet[][ Pisa]{pisa12}, (both
  with $Z=0.004, Y=0.250$) and \citet[][ BaSTI]{basti04} (with
  $Z=0.004, Y=0.251$).}
  \label{fig_comp}
\end{figure*}

\begin{figure*}
\begin{minipage}{0.49\textwidth} 
\resizebox{\hsize}{!}{\includegraphics{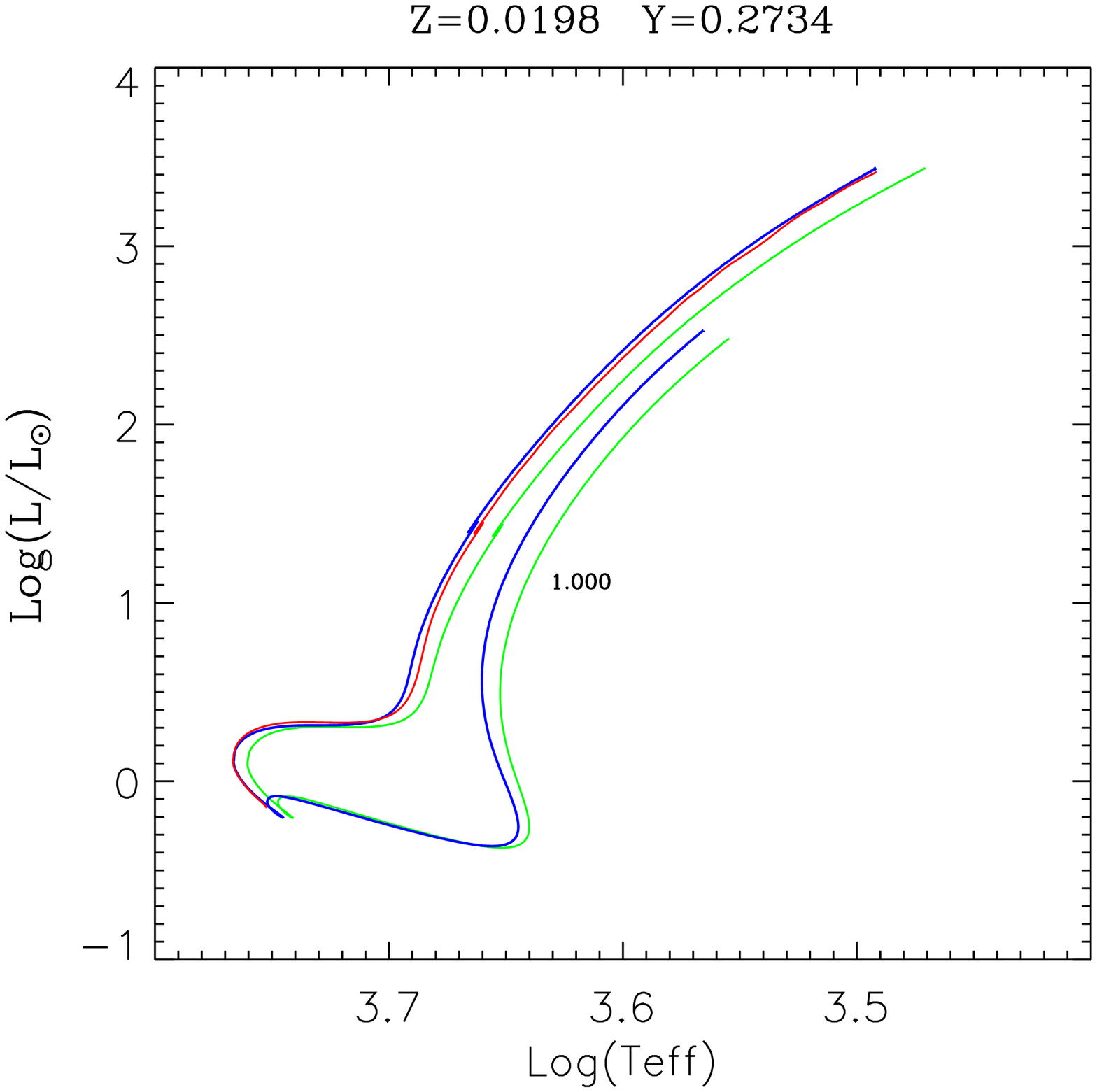}}
\end{minipage}
%
\hfill
\begin{minipage}{0.49\textwidth} 
\resizebox{\hsize}{!}{\includegraphics{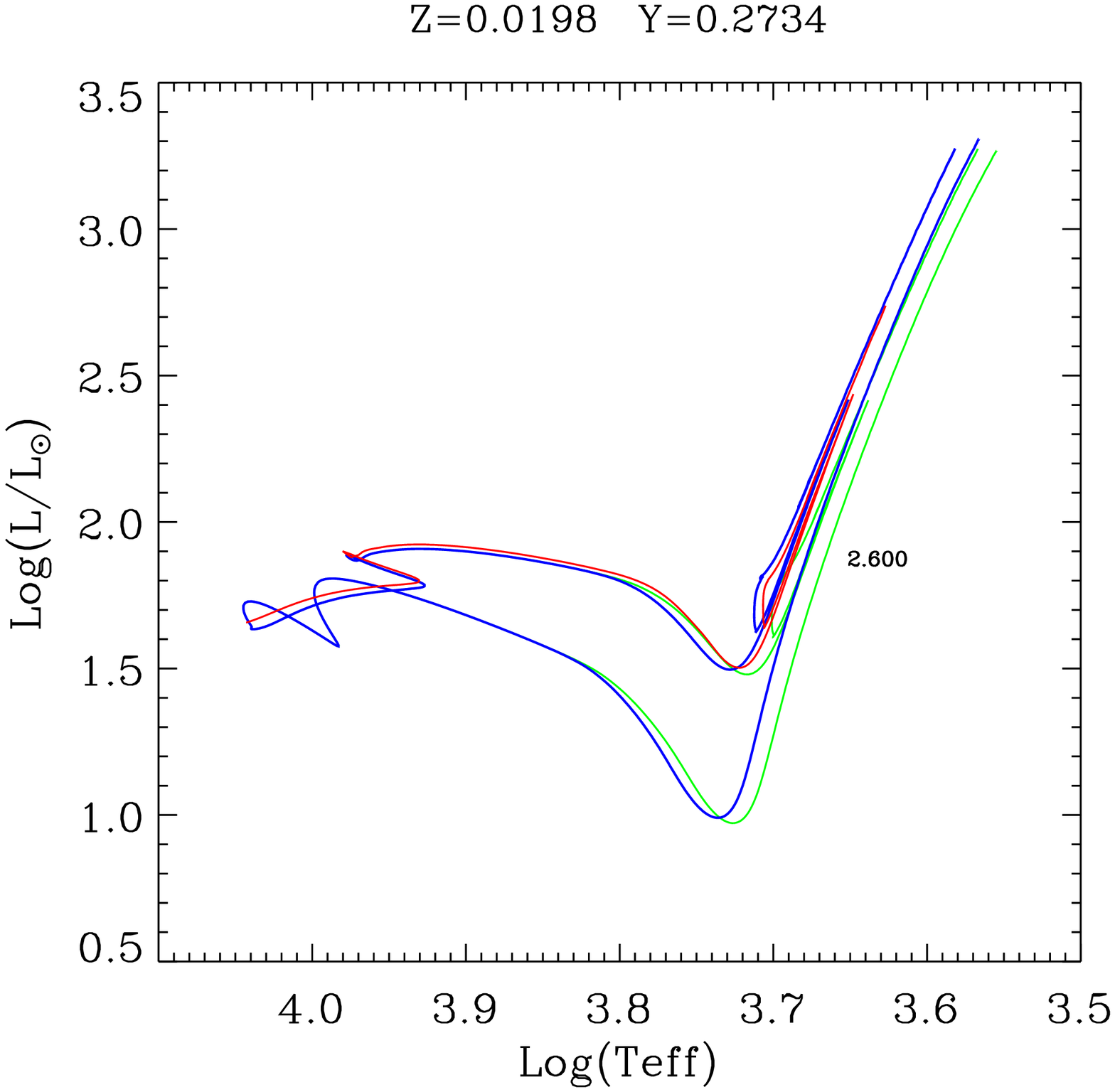}}
\end{minipage}
%
\caption{Comparison with BaSTI tracks \citep{basti04} in the HR
  diagram. The two panels show tracks for masses of 1.0 and
  2.6~\Msun\ obtained from the BaSTI (red lines) database for the same
  fixed solar-scaled abundance of $Z=0.0198,Y=0.2734$ (based on
  \citealt{GrevesseSauval_98}) and $\alpha_{\rm MLT}=1.913$. The blue
  lines are PARSEC scaled-solar tracks obtained for the same
  composition and $\alpha_{\rm MLT}$ value as in BaSTI, whereas the
  green lines are the same but using the PARSEC calibration of the
  mixing length parameter, i.e.\ $\alpha_{\rm MLT}=1.74$.}
\label{fig_compbasti}
\end{figure*}

\subsection{Comparison with other databases}
\label{sec_compbasti}

There are several large databases of stellar evolutionary tracks
available in the literature. All them differ in several important
aspects, including the adopted solar chemical composition and
calibration, the sources of opacities and EOS, the coverage of
evolutionary phases, the extension in the metal-helium content plane,
etc. Comparing all these tracks is not among our goals.  We limit
ourselves to a few comparisons that point to the important
similarities and differences between some recent libraries.

Figure~\ref{fig_comp} compares the He core mass and the luminosity at
the He-flash (or, equivalently, at the tip of the RGB), with those
provided in the Dartmouth \citep{dartmouth08}, Pisa \citep{pisa12} and
BaSTI \citep{basti04} databases. It is evident that, although the
initial helium content in not exactly the same among these tracks, the
PARSEC models agree very well with the most recent Pisa ones, and
follow the same trend as the Dartmouth ones -- with just small
offsets, of a few 0.001~\Msun\ in the He core mass, and a few 0.01~dex
in luminosity. This general agreement is very encouraging. It is also
interesting to note that, although presenting offsets of similar
magnitude, the BaSTI tracks have a different behaviour, with slightly
smaller core masses being followed by slightly larger luminosities at
the tip of the RGB.

As an illustration of the possible differences in the HR diagrams,
Fig.~\ref{fig_compbasti} compares PARSEC with BaSTI tracks
\citep{basti04}, using always the same chemical composition of
$Z=0.0198,Y=0.2734$. This is the initial solar composition in BaSTI,
obtained from \citet{GrevesseSauval_98} and their calibration of the
solar model, which results in a mixing length parameter of
$\alpha_{\rm MLT}=1.913$. The straight computation of PARSEC stellar
evolutionary tracks with the same $Z$, $Y$, and $\alpha_{\rm MLT}$
results in tracks very similar to BaSTI, as can be appreciated
comparing the red (BaSTI) with the blue lines (PARSEC).  However, the
bulk of present PARSEC tracks are calculated with our own
solar-calibrated $\alpha_{\rm MLT}=1.74$.  PARSEC tracks with
$\alpha_{\rm MLT}=1.74$ are presented in green.  They appear
significantly cooler than the BaSTI tracks, especially at the RGB.
This significant difference results mainly from the lower
metallicities assumed for the solar model in PARSEC, and the
calibration of $\alpha_{\rm MLT}$.

This example illustrates how risky is the straight comparison between
tracks computed with a different solar composition and calibration.

\subsection{Data tables}
\label{sec_tabletrack}

The data tables for the present evolutionary tracks are available in
electronic format only through our websites at
OAPD\footnote{http://stev.oapd.inaf.it/parsec\_v1.0}.  For each
evolutionary track, the corresponding data file presents at least 50
columns with the run of many quantities of interest along the track,
as described in a suitable {\sl ReadMe} file.

\section{Isochrones}
\label{sec_isochrones}

For the sake of completeness we present here, together with the set of
evolutionary tracks, also the corresponding isochrones.  These
isochrones do not contain neither the TP-AGB nor the post-AGB phases.
These are discussed in accompanying papers dealing with the extension
of the tracks into these advanced phases (e.g. Marigo et al., in
preparation), and their calibration (basically in terms of third
dredge-up and mass loss efficiencies) on the base of a large set of
observational data.  Although we can ensure a quick initial
calibration of the TP-AGB phase following the procedure defined by
\citet{MarigoGirardi07}, a thoroughful calibration may require a
significant time before completion. We thus focus here our attention
on the less advanced phases of stellar evolution which are,
nevertheless, important thanks to the huge amount of observational
data they can help to interpret.

Following \citet{Bertelli_etal94}, isochrones are computed by first
dividing the evolutionary tracks in suitable homogeneous evolutionary
phases.  The isochrones are then constructed by interpolating points
along missing stellar tracks, between the corresponding phases of the
calculated grid masses. The points that separate the selected
evolutionary phases along the tracks are automatically recognized by a
suitable algorithm that takes into account both the morphology in the
HR diagram and the interior physics.  We store the full track
information for the purpose of constructing isochrones even in
non-conventional diagrams, like for example in the surface composition
of some particular elements.

\begin{figure*}
\begin{minipage}{0.48\textwidth} 
\resizebox{\hsize}{!}{\includegraphics{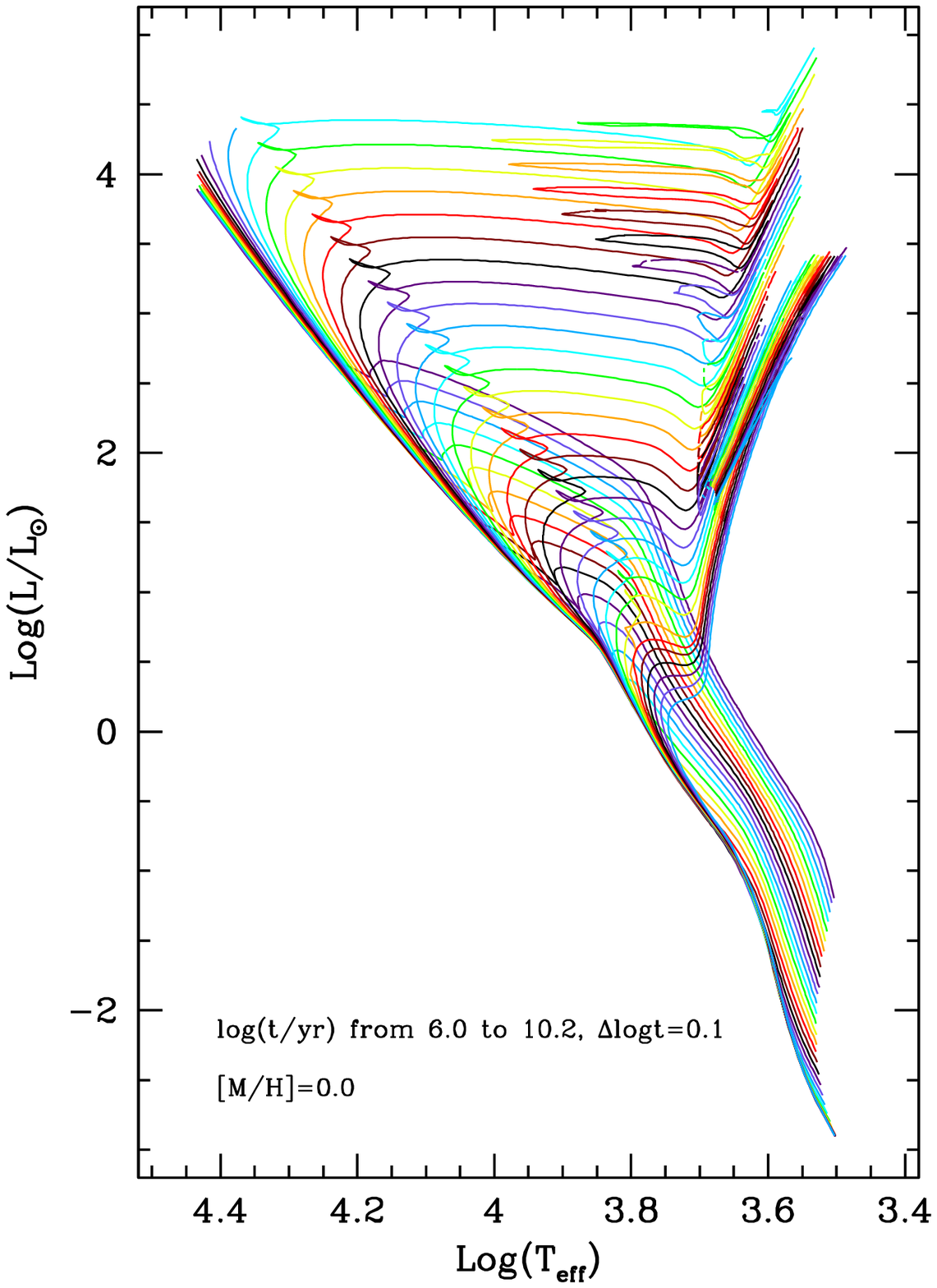}}
\end{minipage}
\hfill
\begin{minipage}{0.48\textwidth} 
\resizebox{\hsize}{!}{\includegraphics{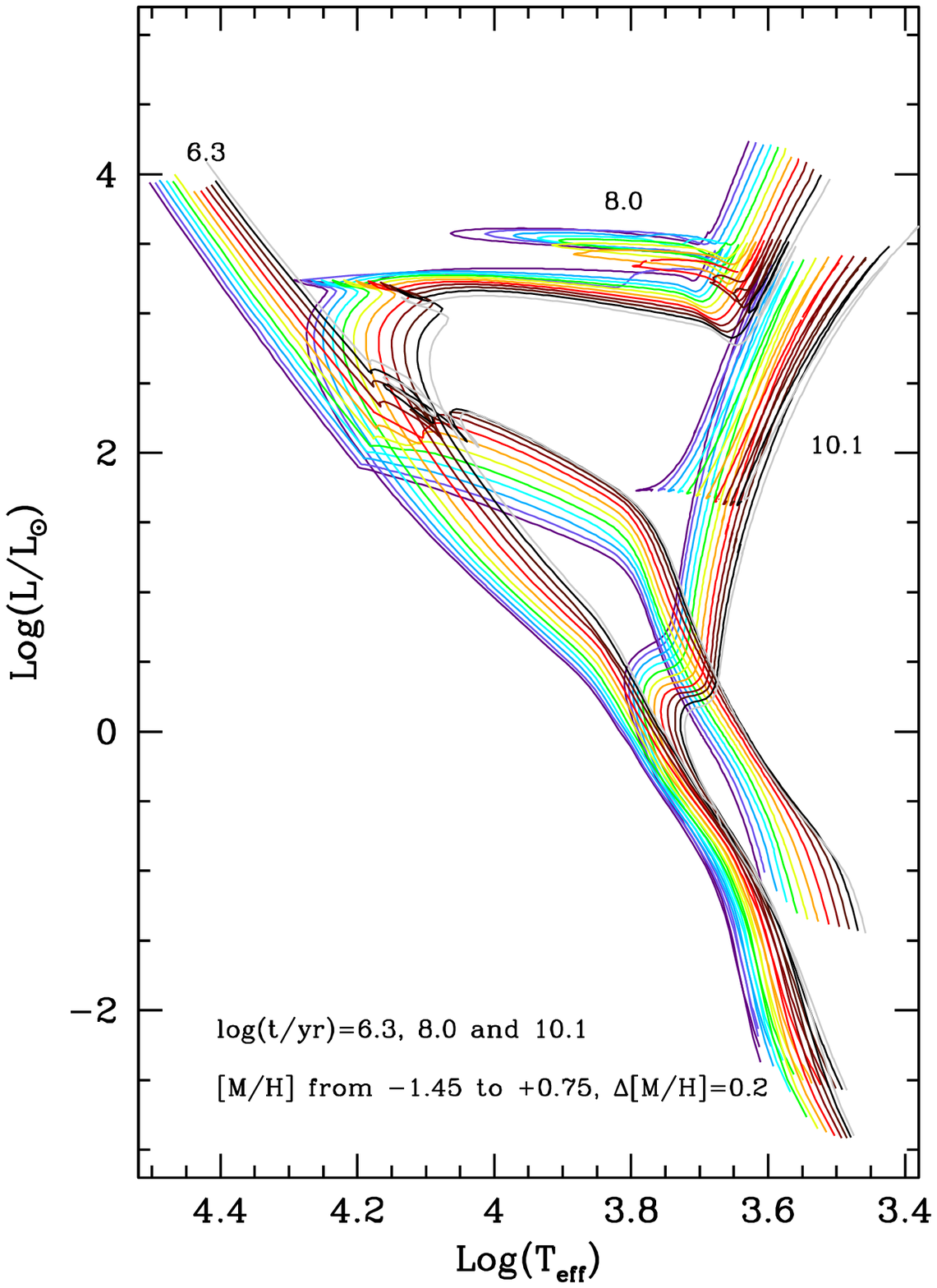}}
\end{minipage}
\caption{Examples of theoretical isochrones in the HR diagram.  {\bf
    Left panel:} A sequence of solar-metallicity isochrones for ages
  going from $\logt=6$ to $10.2$ at equally spaced intervals of
  $\Delta\log t=0.1$.  {\bf Right panel:} Three sequences of
  isochrones for a fixed age and at varying metallicty. The ages are
  $\logt=6.3$, 8.0 and 10.1. The sequence of metallicities goes from
  $[{\rm M/H}]=-1.45$ to $+0.75$ at equally spaced intervals of
  $\Delta[{\rm M/H}]=0.2$. In all cases, the mass-loss parameter is
  assumed to be $\eta=0.2$, and the sequences are completed down to
  0.1~\Msun. Notice the presence of the PMS phase in the youngest
  isochrones.}
\label{fig_isocrone}
\end{figure*}

Examples of isochrones in the HR diagram are shown in
Fig.~\ref{fig_isocrone}, for the sets with scaled-solar chemical
composition that follow the relation $Z=0.2485+1.78\,Z$. For these
sets, isochrones can already be built for any age from $\sim\!1$~Myr
to at least 20~Gyr (limited to the mass interval $0.1\le
M\le12$~\Msun), and for any metallicity in the range $0.0005\le
Z\le0.07$ (corresponding to $-1.49\le{\rm [M/H]}\le+0.78$). We also
note the presence of the PMS in the isochrones with the youngest ages.
Mass loss by stellar winds has been considered only during the RGB of
low-mass stars, using the empirical formula by \citet{reimers}
multiplied by an efficiency factor $\eta$. The latter is set as
$\eta=0.2$, which is representative of the modest values recently
determined by \citet{Miglio_etal12}, based on the asteroseismic data
for two star clusters in the {\em Kepler} fields.

The initial point of each isochrone is the 0.1~\Msun\ model in the
lower main sequence. The terminal stage of the isochrones is either
the beginning of the TP-AGB for low- and intermediate-mass stars
($M\le M_{\rm IM}$, ages $\ga10^8$~yr), or C-ignition for more massive
stars.

Theoretical luminosities and effective temperatures along the
isochrones are translated to magnitudes and colors using extensive
tabulations of bolometric corrections and colors, as detailed in
\citet{Girardi_etal02}, \citet{Marigo_etal08}, \citet{Girardi_etal08},
and \citet{Rubele_etal12}. The tabulations were obtained from
convolving the spectral energy distributions contained in the
libraries of stellar spectra of \citet{CastelliKurucz03},
\citet{Allard_etal00, Allard_etal01}, and \citet{Fluks_etal94} with
the response function of several medium- and broad-band filters. The
reference spectrum for Vegamag systems is taken from \citet{Bohlin07}.

The bulk of stars modelled in optical filters, for \Teff\ between
50\,000 and $\sim\!4000$~K, have their colors based on the
\citet{CastelliKurucz03} ATLAS9 models. \citet{Casagrande_etal10} have
recently and accurately revised the temperature scale of nearby stars
via the infrared flux method. It is extremely reassuring to note that,
for the Sun, the color differences between their calibration and
\citet{CastelliKurucz03} models are in general only $\simeq\!0.02$~mag
(which is also comparable to the random errors in the calibration; see
their table~7), all the way from the $B$ to the $H$ band.

The isochrones are initially made available through the CMD web
interface at OAPD\footnote{http://stev.oapd.inaf.it/cmd}. It allows
isochrones to be retrieved for any given choice of age and initial
metallicity, within the range of the calculated tracks. Output tables
include the most relevant stellar parameters along the isochrones
(e.g. $L$, $\Teff$, $g$, radius, evolutionary phase), and the absolute
magnitudes in the photometric system of choice.  Isochrones are now
provided in more than 30 photometric systems corresponding to major
astronomical facilities and surveys. However, the list of photometric
systems is being continuously expanded, and updated in response to
changes into filter transmission curves and zero-points.  We
anticipate that the same web interface is being restructured in order
to deal with the entire variety of stellar parameters stored in the
PARSEC evolutionary tracks.

\section{Discussion and concluding remarks}
\label{sec_conclu}

\begin{figure}
\resizebox{\hsize}{!}{\includegraphics{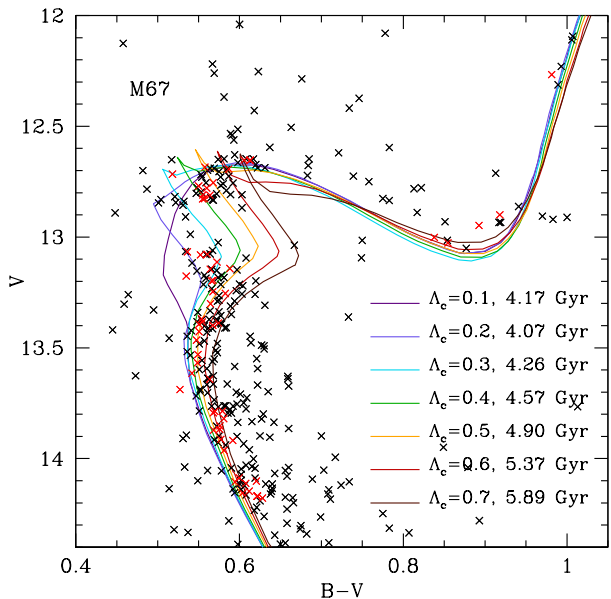}}
\caption{PARSEC isochrones overlaid in the CMD of the intermediate-age
  star cluster M\,67. The distance modulus and colour excess were
  fixed as $V-M_V=9.75$ and $E_{B\!-\!V}=0.03$, as shown in
  Fig.~\ref{fig_clusters} next. The $BV$ photometry is taken from
  \citet[][black points]{Montgomery_etal93}. Red dots mark the
  ``high-probability single stars'' identified by \citet{Sandquist04}.
  The isochrones are for $Z=0.014$, and for several values of
  overshooting parameter $\Lambda_{\rm c}$. The ages are selected so
  as to reasonably fit the position of the subgiant branch, and
  especially its lower boundary which coincides with the locus of
  bona-fide single stars. Notice that the isochrones with
  $\Lambda_{\rm c}<0.4$ and $\Lambda_{\rm c}>0.6$ fail in reproducing
  the position and shape of the turn-off region.}
\label{fig_m67new}
\end{figure}

We have discussed the main features of the new code PARSEC that will
be used to update and extend the Padova database of stellar
evolutionary tracks and isochrones.  In PARSEC we have modified and
updated all the major input physics, including the equation of state,
the opacities, the nuclear reaction rates and nuclear network, and the
inclusion of microscopic diffusion.

With the new code we have computed several sets of stellar
evolutionary tracks that differ from the previous ones in a few
additional aspects. (1) They are based on a new reference solar
composition from \citet{Caffau_etal11}. The present code however
allows the quick computation of tracks with modified metal distributions
(e.g. $\alpha$ enhancement or depletion), fully taking into account
the changes implied to tables of opacities and equation of state. (2)
We consider the evolution along the PMS phase. (3) A few changes have
been introduced on the way the convection is treated.  From these
tracks we obtain preliminary isochrones up to the beginning of the
TP-AGB.

\begin{figure*}
\resizebox{0.47\hsize}{!}{\includegraphics{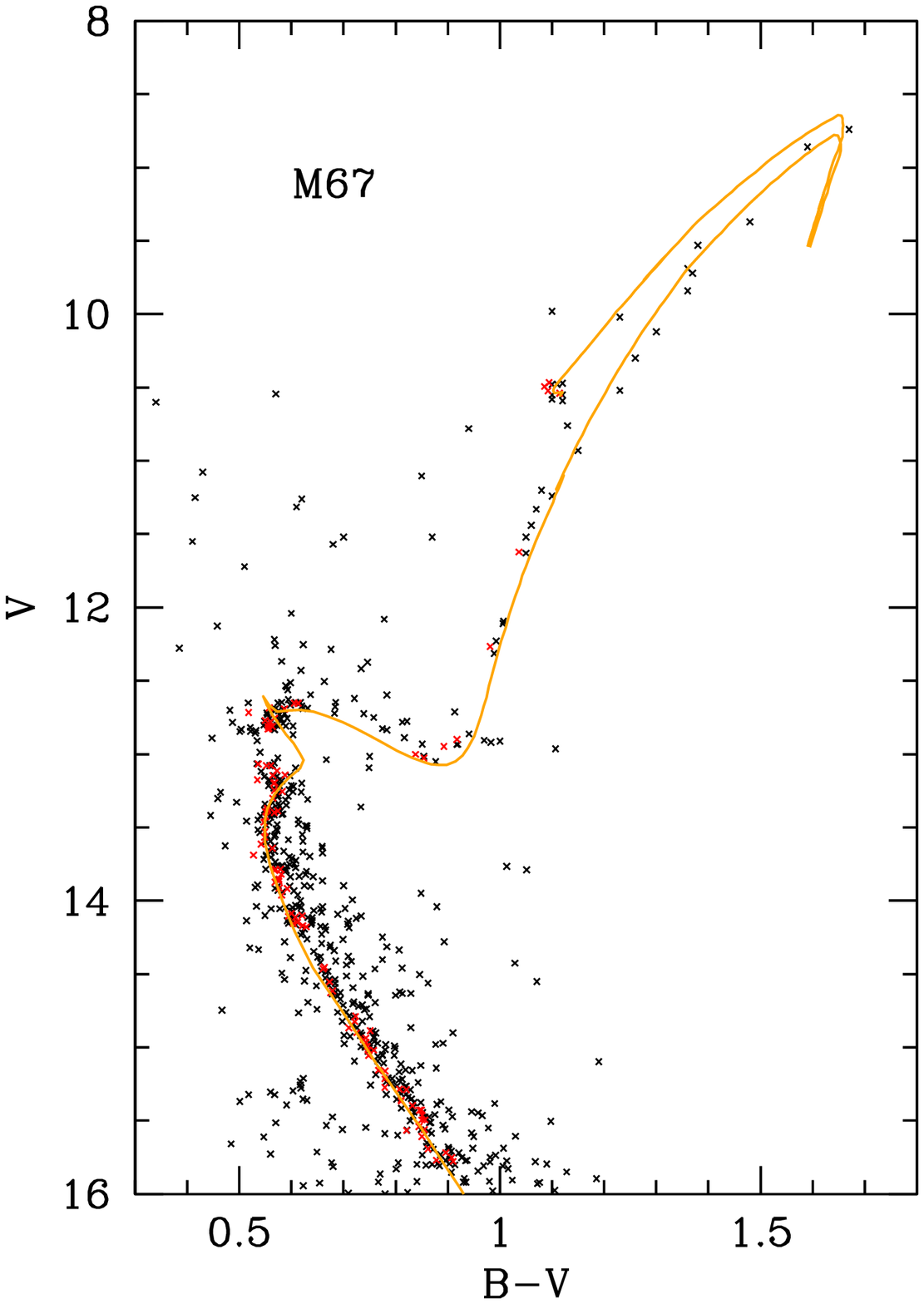}}
\hfill
\resizebox{0.47\hsize}{!}{\includegraphics{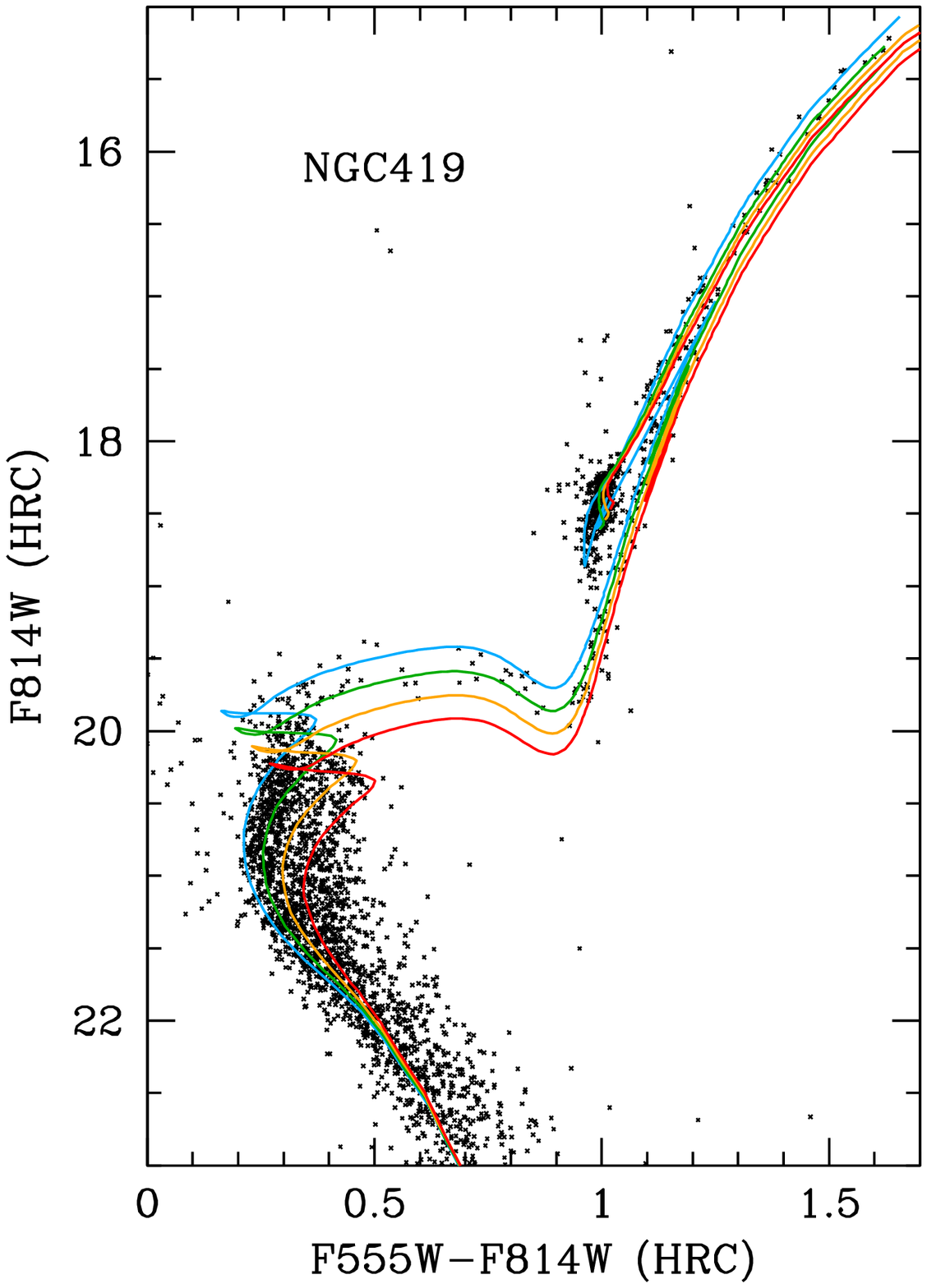}}
\caption{PARSEC isochrones overlaid in the CMDs of two
  intermediate-age star clusters. In both cases, the distance modulus
  and reddening were fixed ``by eye'' so as to reproduce the position
  of the red clump. {\bf Left panel:} The Galactic open cluster M\,67.
  The data are the same as in Fig.~\ref{fig_m67new}.  The isochrones
  are for $Z=0.014$ and $\log(t/{\rm yr})=9.6$, distance modulus of
  $V-M_V=9.75$ and color excess $E_{B\!-\!V}=0.03$.  {\bf Right
    panel:} The core of the SMC cluster NGC~419. The photometry comes
  from archival HST images obtained with the High Resolution Channel
  of the Advanced Camera for Surveys onboard HST (GO-10396, PI: J.S.
  Gallagher), and reduced as described in \citet{Girardi_etal09}. In
  order to reduce the presence of blends and spurious objects, we plot
  only those stars with a sharpness of $({\rm sharp}_{\rm F555W}^2+{\rm
    sharp}_{\rm F814W}^2)^{0.5}<0.1$.  The isochrones are for
  $Z=0.004$, ages $\log(t/{\rm yr})=9.15, 9.20, 9.25, 9.30$ (from blue
  to red), distance modulus of ${\rm F814W}-M_{\mathrm F814W}=18.96$
  and color excess $E_{{\mathrm F555W}\!-\!{\mathrm F814W}}=0.09$.}
\label{fig_clusters}
\end{figure*}

Examples of the quality of the new isochrones are shown in
Figs.~\ref{fig_m67new} and \ref{fig_clusters} where we present
preliminary fits of the colour magnitude diagrams of the Galactic open
cluster M\,67 and the SMC cluster NGC~419.

M\,67 is the best example of an intermediate-age cluster in which the
turn-off region presents the clear signatures of well-developed
convective cores, in particular the sizeable gap in the concentration
of stars at the termination of the main sequence, which is followed by
the rapid contraction of the previously convective cores, soon after
H-exhaustion at the centre.  This cluster indicated the need for a
quick increase of overshooting efficiency with mass, as advanced in
Sect.~\ref{sec_conv}. Indeed, Fig.~\ref{fig_m67new} shows a series of
isochrones obtained from PARSEC tracks with overshooting parameter
going from $\Lambda_{\rm c}=0.1$ to 0.7, for the near-solar
composition of $(Z=0.014, Y=0.273)$ ($[{\rm M/H}]=-0.02$). In all
cases, the isochrone age was fixed so as to produce about the same
description of the subgiant phase, along the line that joins the
bona-finde single stars located between the reddest termination of the
turn-off region and the lower extremity of the fisrt-ascent RGB. The
distance and reddening of the isochrones is fixed so as to
satisfactorily reproduce the mean position of the RGB, the lower main
sequence, and the red clump of He-burning stars, as can be appreciated
in the left panel of Fig.~\ref{fig_clusters}\footnote{We recall that
  for the typical masses in these isochrones, the red clump position
  does not vary with the overshooting efficiency, since the core mass
  at the RGB tip is practically insensitive to this parameter.}.
Therefore, all these isochrones differ mainly in their description of
the turn-off region. The figure reveals that isochrones with
$\Lambda_{\rm c}<0.4$ present a too blue turn-off, compared to the
locus of bona-fide single stars in M67. Conversely, isochrones with
$\Lambda_{\rm c}>0.5$ have clearly too red a turn-off. This comparison
supports our initial choice for the value of overshooting efficiency
for masses $M>M_{\rm 02}$, as described in Sect.~\ref{sec_conv}. More
detailed work is underway, considering the full range of parameters
(distances, reddening, metallicity, binary frequency, etc.), as well
as the detailed predictions about the occupation probability of
different sections of the isochrones.

We also note that M\,67 has been object of many recent analyses aimed
at deriving accurate determinations of its age, and constraints on its
chemical composition and convective efficiency \citep[see
e.g.][]{Michaud_etal04, VandenBerg04, VandenBerg_etal07, Magic_etal10,
  Mowlavi_etal12}. These works were intentionally focused in the
turn-off region of the CMD -- and especially on its hook-like feature
-- and did not attempt to fit the RGB and red clump altogether with
the same isochrone, as we do in Fig.~\ref{fig_clusters}. Our choice of
a ``global isochrone fitting'' is based on a very simple
consideration: our evolutionary tracks are mostly aimed at improving
evolutionary population synthesis of resolved and unresolved galaxies,
for which most observations include very well sampled giant branches.
Therefore, a good reproduction of the evolved phases in the CMD is as
important in our case, as is the good reproduction of the turn-off
region in those more focused works, where the main target is the
calibration of physical processes and abundances inside stars.

Another interesting object for a preliminary check of the efficiency
of overshooting is the intermediate-age SMC cluster NGC~419.  Its CMD
(right panel in Fig.~\ref{fig_clusters}) presents clear evidences of
multiple populations, namely a broad turn-off and a dual red clump
\citep{Girardi_etal09}.  The turn-off masses are around the $M_{\rm
  HeF}$ limit between intermediate- and low-mass stars, as evidenced
by the double structure of the red clump.  For a cluster of this kind,
reproducing the shape of both the turn-off and red clump regions is
possible only if the isochrones have about the right amount of
overshooting, as discussed in \citet{Girardi_etal09}. As illustrated
in Fig.~\ref{fig_clusters}, the default PARSEC isochrones (with
$\Lambda_{\rm c}>0.5$ in the mass range of interest for NGC~419) with
ages spanning from 1.4 to 2.0 Gyr are able to reproduce both the
``golf-club'' shape of the turn-off region, and the dual structure of
the red clump -- where the secondary, fainter red clump is explained
by the youngest isochrone.

For the moment, our goal is just to call attention to the correct
reproduction of the main CMD features, including the shape of the
turn-off(s), the subgiant branch, the RGB slope, and the red clump(s)
in these two clusters.  We anticipate that a detailed quantitative
analysis of these and other similar clusters is being performed to
calibrate the efficiency of overshoot in this mass interval (Rubele et
al., in prep.). Younger stellar populations will be considered by
Rosenfield et al. (in prep.).  The evolution along the TP-AGB, and the
corresponding isochrones, will be presented in an accompanying paper
(Marigo et al., in prep.).


\section*{Acknowledgements}

We thank the anonymous referee for the many suggestions that helped us
to improve the final version of this paper, and M. Barbieri, G.
Bertelli, A. Miglio and P. Rosenfield for helpful discussions.  We
acknowledge financial support from contract ASI-INAF I/009/10/0.
A.B. acknowledges financial support from MIUR 2009.

%

%
\label{lastpage}
\end{document}